\documentclass[pra,showpacs,twocolumn,aps]{revtex4-1}

\usepackage{amsmath}
\usepackage{amssymb}
\usepackage{latexsym}
\usepackage{amsfonts}
\usepackage{epsfig}
\usepackage{psfrag}
\usepackage{graphicx}
\usepackage{color}

\newcommand{\llb}{\langle\langle}
\newcommand{\rr}{\rangle\rangle}

\begin{document}

\title{Hierarchy of double-time correlations} 

\author{Friedemann Queisser and Ralf~Sch\"utzhold}

\affiliation{Fakult\"at f\"ur Physik,
Universit\"at Duisburg-Essen, Lotharstra{\ss}e 1, Duisburg 47057, Germany,}

\affiliation{Helmholtz-Zentrum Dresden-Rossendorf, 
Bautzner Landstra{\ss}e 400, 01328 Dresden, Germany,}

\affiliation{Institut f\"ur Theoretische Physik, 
Technische Universit\"at Dresden, 01062 Dresden, Germany.}

\date{\today}

\begin{abstract}
The hierarchy of correlations is an analytical approximation method which allows us to 
study non-equilibrium phenomena in strongly interacting quantum many-body systems on 
lattices in higher dimensions. 
So far, this method was restricted to equal-time correlators 
$\langle\hat A_\mu(t)\hat B_\nu(t)\rangle$. 
In this work, we generalize this method to double-time correlators 
$\langle\hat A_\mu(t)\hat B_\nu(t')\rangle$, which allows us to 
study effective light cones and Green functions and to 
incorporate finite initial temperatures.
\end{abstract}

\maketitle

\section{Introduction}

The physics of strongly interacting quantum many-body systems 
is, despite decades of research, far from being fully understood. 
Apart from a few exactly solvable models, even their ground state properties 
are a subject of ongoing discussions, see, e.g., \cite{L04,YJNIHY03}.
Even more challenging is the investigation of non-equilibrium 
properties of those interacting many-body systems.

In the limit of infinite dimensions, the Dynamical Mean-field Theory (DMFT) 
has been successfully applied to various problems such as the Mott-Hubbard 
transition \cite{RZK92,RKZ94,GK92,GKKR96,AGPTW10,AGPTW11} or the non-equilibrium dynamics in Mott insulators \cite{EKW09,ATE14,EKW10,WTE12,CWE14,FT06} 
by mapping the system to an effective single site (i.e., zero-dimensional) 
problem. 

Furthermore, one-dimensional strongly interacting systems have been studied 
by employing exact diagonalization, see, e.g., \cite{K16,LHS88,KLA07,KTEG08,RB04}, 
time-dependent Density Matrix Renomalization Group variational techniques (t-DMRG),  
see, e.g., \cite{DKSV04,S11,WF04}, or Jordan-Wigner transformations \cite{JW28,LP75,BO01,BKOSV04}. 
Moreover, exact analytical solutions were used to study non-equilibrium 
dynamics \cite{D05} and ground state properties.
However, the findings for one-dimensional systems cannot be 
easily transferred to higher dimensions.
For example, thermalization in higher dimensions is very different 
from thermalization in one dimension since for the latter
the energy re-distribution cannot occur via two-body collisions
(due to energy and momenta conservation).

For strongly interacting systems in higher dimensions, 
the methods described above run into difficulties.
%
%
For example, the generalization of t-DMRG to higher dimensions 
(such as tensor networks) is limited by the exponential scaling 
with the system width \cite{CV09}, especially if correlations spread 
across the system. 
%
%
For a very similar reason (exponential scaling of Hilbert space), 
the method of exact diagonalization is limited to 
small system sizes \cite{K16}.
On the other hand, the DMFT is only well understood in the limit of infinite 
dimensions and, due to the mapping to an effective single-site problem, does 
not capture energy and momentum transfer of quasi-particles or long-range correlations 
\cite{WEFWBH18}.

In order to bridge this gap, 
we established over the last years a perturbative hierarchical method which 
is valid for large coordination numbers $Z\gg1$.
This method allows for a systematic study of non-equilibrium properties in 
strongly interacting systems in large (but finite) dimensions \cite{NS10,QNS12,QKNS14,KNQS14,NQS14,NQS16,QS19,QS19a}.

The hierarchical expansion is based on a controlled expansion 
of the $n$-point reduced density matrices into correlated parts.
At zeroth order in our expansion, we have the single-site density matrix,
\begin{equation}
\hat{\rho}_\mu=\mathrm{Tr}_{\not{\mu}}\{\hat\rho\}=
\mathcal{O}(Z^0)\,,
\end{equation}
where $\mathrm{Tr}_{\not{\mu}}$ denotes the trace over all lattice sites 
but $\mu$. 
The correlated part of the two-site density matrix, 
\begin{equation}
\hat{\rho}^\mathrm{corr}_{\mu\nu}
=\mathrm{Tr}_{\not{\mu}\not{\nu}}\{\hat\rho\}-\hat{\rho}_\mu \hat{\rho}_\nu
\,, 
\end{equation}
is of order $\mathcal{O}(1/Z)$, the three-point correlator 
\begin{equation}
\hat{\rho}^\mathrm{corr}_{\mu\nu\lambda}
=\mathrm{Tr}_{\not{\mu}\not{\nu}\not{\lambda}}
\{\hat\rho\}-\hat{\rho}^\mathrm{corr}_{\mu\nu} 
\hat{\rho}_\lambda-\hat{\rho}_{\mu\lambda}^\mathrm{corr} 
\hat{\rho}_\nu -\hat{\rho}_{\nu\lambda}^\mathrm{corr} 
\hat{\rho}_\lambda-\hat{\rho}_\mu \hat{\rho}_\nu \hat{\rho}_\lambda 
\end{equation}
is of order $\mathcal{O}(1/Z^2)$, and so on.
The time-evolution of the $n$-point correlations is derived 
directly from the von Neumann equations.
With this expansion, we were able to study quenches 
across phase boundaries \cite{NS10}, ground state properties and non-equilibrium 
dynamics of quantum correlations in the bosonic and fermionic 
Hubbard model \cite{QKNS14,KNQS14}.
Although the perturbative scheme is set up for $Z\gg1$, we 
found even qualitative agreement with exact diagonalization  in one dimension \cite{KNQS14}. 

The derivation of the hierarchy was based on the real-time 
evolution of the system's density matrix.
In the following, we shall extend our hierarchical 
approach to double-time correlation functions.
As a concrete example, we will apply the method to the Bose-Hubbard model 
in the Mott insulating phase.
In section \ref{H} we derive the equations of motion 
up to order $\mathcal{O}(1/Z)$. 
A general proof of the hierarchy is given in the Appendix.
In section \ref{BHequil}, we show how Green functions and finite 
temperature correlation functions are related to the hierarichal equations.
In section \ref{BHnonequil1} we study the non-equilibrium dynamics at finite 
temperatures and in section \ref{BHnonequil2} we derive the double-time correlation 
functions for a quantum quench of the Bose-Hubbard model.



\section{The Double-Time Hierarchy}\label{H}

The lattice system under consideration is described by the Bose-Hubbard 
Hamiltonian ($\hbar=1$)
\begin{align}\label{bosehubbard}
\hat{H}=-\frac{J(t)}{Z}\sum_{\mu,\nu}T_{\mu\nu}\hat{b}^\dagger_\mu\hat{b}_\nu+\sum_\mu \left[\frac{U}{2}\hat{n}_\mu(\hat{n}_\mu-1)-\mu_0 
\hat{n}_\mu\right]\,, 
\end{align}
where $\hat b_{\mu}^\dagger$ and $\hat b_{\nu}$ are the bosonic creation 
and annihilation operators at the lattice sites $\mu$ and $\nu$, respectively.
The time-dependent hopping rate is denoted by $J(t)$, where we have 
factored out the coordination number $Z$.
The second term describes the on-site repulsion $U$ with the particle 
number operator $\hat n_\mu$ and the chemical potential~$\mu_0$.

The system (\ref{bosehubbard}) can only be solved exactly for a small number of lattice sites. 
For large systems one relies on suitable approximation schemes.
As mentioned in the introduction, we shall employ 
a hierarchy of correlations for large coordination numbers $Z$.
For arbitrary operators $\hat A_\mu(t)$ and $\hat B_\mu(t)$ 
we have (see the Appendix)
\begin{align}
\langle \hat A_\mu(t) \hat B_\mu(t')\rangle&=\mathcal{O}(Z^0)\\
\langle \hat A_\mu(t) \hat B_\nu(t')\rangle^\mathrm{corr}&=\langle \hat A_\mu(t) \hat B_\nu(t')\rangle-\langle \hat A_\mu(t)\rangle\langle \hat B_\nu(t')\rangle\nonumber\\
&=\mathcal{O}(1/Z),\, \mathrm{etc.}
\end{align}
We assume the system to be in the Mott insulating phase
(with integer filling)
where the on-site repulsion $U$ dominates over the hopping rate $J$. 
To order $\mathcal{O}(Z^0)$, the density matrix of the system 
is given by
\begin{align}
\hat{\rho}=\bigotimes_\mu \hat\rho_\mu, \quad
\hat\rho_\mu=\sum_n p_n(t)|n\rangle_\mu \langle n| 
\end{align}
with the on-site probabilities $p_n(t)$.
%
We define the operators $\hat{P}^{n,m}_\mu=|n\rangle_\mu\langle m|$
and find for the on-site expectation values
\begin{align}\label{singlesite}
&i\partial_t\langle \hat{P}^{n,n+1}_\mu(t) \hat{P}^{m+1,m}_\mu(t')\rangle\nonumber\\
&=(U n -\mu_0)\langle \hat{P}^{n,n+1}_\mu(t) \hat{P}^{m+1,m}_\mu(t')\rangle
+\mathcal{O}(1/Z)\,.
\end{align}
The two-site correlation functions with $\mu\neq\nu$
evolve according to
%
%
%
\begin{align}\label{twosite}
&\left(i\partial_t-U n +\mu_0\right)\langle \hat{P}^{n,n+1}_\mu(t) 
\hat{P}^{m+1,m}_\nu(t')\rangle^\mathrm{corr}\nonumber\\
=&-\frac{J(t)}{Z}\sum_{\kappa\neq \mu,\nu}T_{\mu\kappa}\sqrt{n+1}[p_n(t)-p_{n+1}(t)]\nonumber\\
&\times \sum_{l=0}^\infty\sqrt{l+1}\langle \hat{P}^{l,l+1}_\kappa(t)\hat{P}^{m+1,m}_\nu(t')\rangle^\mathrm{corr}\nonumber\\
&-\frac{J(t)}{Z}T_{\mu\nu}\sqrt{n+1}[p_n(t)-p_{n+1}(t)]\nonumber\\
&\times \sum_{l=0}^\infty\sqrt{l+1}\langle \hat{P}^{l,l+1}_\nu(t)\hat{P}^{m+1,m}_\nu(t')\rangle+\mathcal{O}(1/Z^2)\,.
\end{align}
In equation (\ref{twosite}) we employed the hierarchy 
in order to separate
three-point expectation values, 
\begin{align}
&\langle\hat{P}_\mu^{n,n}(t) \hat{P}^{l,l+1}_\kappa(t)
\hat{P}^{m+1,m}_\nu(t')\rangle=\nonumber\\
&p_n(t)\langle\hat{P}^{l,l+1}_\kappa(t)
\hat{P}^{m+1,m}_\nu(t')\rangle^\mathrm{corr}+\mathcal{O}(1/Z^2)\,, \mathrm{etc.} 
\end{align}
In the following we shall study only the 
evolution up to order $\mathcal{O}(1/Z)$.
Thus we limit our considerations to the linear evolution 
of quasi-particle excitations.
All results in this work will be an immediate 
consequence of the evolution equations (\ref{singlesite}) and (\ref{twosite}).

Note that the hierarchical expansion can be extended to 
arbitrary high orders in $1/Z$, e.g., the interaction 
among the quasi-particle excitation can be taken into account \cite{QS19a}.
The proof of whole $1/Z$-hierarchy for double-time 
correlation functions involving arbitrary $n$-point 
correlations is given in the appendix \ref{PH}.

\section{Double-time expectation values in equilibrium}\label{BHequil}

The hierarchical set of equations for two-time expectation values 
induces a hierarchy for double-time Green's functions.
In the following we will use them for the computation 
of equilibrium correlation functions in the Bose-Hubbard model at finite temperatures.
A thermal expectation value of an operator $\hat O $ is defined as
\begin{eqnarray}
\langle \hat O\rangle_\mathrm{therm}=\frac{\mathrm{tr}(\hat{O}e^{-\beta\hat{H}})}{\mathrm{tr}
(e^{-\beta\hat{H}})}\,,
\end{eqnarray}
where $\beta=1/k_b T$ is the inverse temperature.
The retarded and advanced Green's functions for 
operators $\hat{A}_\mu(t)$ and $\hat{B}_\nu(t')$  are given by \cite{Z60}
\begin{eqnarray}\label{green}
G^r_{A_\mu,B_\nu}(t,t')&=&\llb \hat{A}_\mu(t);\hat{B}_\nu(t')\rr^{r}\nonumber\\
&=&-i \Theta(t-t')
\langle[\hat{A}_\mu(t),\hat{B}_\nu(t')]\rangle
\end{eqnarray}
and 
\begin{eqnarray}
G^a_{A_\mu,B_\nu}(t,t')&=&\llb \hat{A}_\mu(t);\hat{B}_\nu(t')\rr^{a}\nonumber\\
& =&i \Theta(t'-t)
\langle[\hat{A}_\mu(t),\hat{B}_\nu(t')]\rangle\,.
\end{eqnarray}
Both Green functions, although they are subject to different boundary 
conditions, obey the same differential equation.
We choose $\hat{A}_\mu=\hat{P}^{n,n+1}_\mu$ and $\hat{B}_\mu=\hat{P}^{m+1,m}_\mu$
and find from equations (\ref{singlesite}) and (\ref{green}) the relation
\begin{align}
&i\partial_t\llb \hat{P}^{n,n+1}_\mu(t); \hat{P}^{m+1,m}_\mu(t')\rr
 =\delta(t)\delta_{m,n}(p_n-p_{n+1})\nonumber\\
&+(Un-\mu_0)\llb \hat{P}^{n,n+1}_\mu(t); \hat{P}^{m+1,m}_\mu(t')\rr+\mathcal{O}(1/Z)\,.
\end{align}
The on-site probabilities $p_n$ are time-independent since we evaluate 
all quantities in a thermal equilibrium state
and the hopping rate $J$ is assumed to be time-independent.
An analogous equation for $\mu\neq\nu$ follows from equation (\ref{twosite}). 
Via a Fourier transform w.r.t. space and time, we can turn
the differential equations for the Green functions into  a set of algebraic equations,
\begin{align}
&(\omega-U n+\mu_0 )\llb \hat{P}^{n,n+1}; \hat{P}^{m+1,m}\rr_\omega=
\frac{\delta_{m,n}}{2\pi}(p_n-p_{n-1})\,,\label{onsitegreen}\\
&(\omega-U n+\mu_0 )\llb \hat{P}^{n,n+1}; \hat{P}^{m+1,m}\rr_{\mathbf{k},\omega}=\nonumber\\
&-JT_\mathbf{k}\sqrt{n+1}(p_n-p_{n+1})\sum_{l=0}^\infty \sqrt{l+1}\nonumber\\
&\times\left[\llb \hat{P}^{l,l+1}; \hat{P}^{m+1,m}\rr_{\mathbf{k},\omega}-\llb \hat{P}^{l,l+1}; \hat{P}^{m+1,m}\rr_{\omega}\right]\label{twositegreen}\,.
\end{align}

Using the spectral decomposition of operators, it can be shown that 
thermal expectation-value of double-time correlation functions and 
the Green's functions are related by \cite{Z60}
\begin{align}\label{zub}
\langle 
\hat{B}(t)\hat{A}(t')\rangle_\mathrm{therm}=&i\lim_{\epsilon\rightarrow0^+}\int_{-\infty}^\infty 
d\omega\frac{e^{i \omega (t-t')}}{e^{\beta \omega}-1}\nonumber \\
\times&\left[\llb \hat{A};\hat{B}\rr_{\omega+i\epsilon}-\llb 
\hat{A};\hat{B}\rr_{\omega-i\epsilon} \right]\,.
\end{align}
The exact relation (\ref{zub}) permits the calculation of time-independent 
expectation values ($t=t'$) or double-time expectation values 
($t\neq t'$) at finite temperatures.
From equation (\ref{onsitegreen}) we obtain the probabilities for $n$ bosons on a lattice site,
\begin{align}
p_n&=\frac{e^{-\beta\left(\frac{U}{2}n(n-1)-\mu_0 n\right)}}{\sum_{m=0}^\infty 
e^{-\beta\left(\frac{U}{2}m(m-1)-\mu_0 m\right)}}\,.\label{thermalpn}
\end{align}

From equation (\ref{twositegreen}) we find after some algebra the two-site correlator in order $1/Z$,
\begin{align}\label{corrbb}
&\langle \hat{b}^\dagger_\mu(t)\hat{b}_\nu(t')\rangle_\mathrm{therm}=
\frac{\mathrm{Tr}\left(e^{-\beta\hat{H}}\hat{b}^\dagger_\mu(t)\hat{b}_\nu(t')\right)}
{\mathrm{Tr}\left(e^{-\beta\hat{H}}\right)}\nonumber\\
=&\frac{1}{ \pi N}\int d\omega\, \mathcal{P}\frac{1}{e^{\beta \omega}-1}\sum_\mathbf{k}\mathrm{Im}\left(\frac{JT_\mathbf{k}G(\omega+i\epsilon)^2}
{1+JT_\mathbf{k}G(\omega+i\epsilon)}\right)\nonumber\\
&\times e^{i\mathbf{k}\cdot (\mathbf{x}_\mu-\mathbf{x}_\nu)+i\omega(t-t')}\,,
\end{align}
where we introduced the thermal Green function \cite{NQS16}
\begin{align}
G(\omega)&=\sum_{n=0}^\infty \frac{(n+1)(p_n-p_{n+1})}{\omega-Un+\mu_0}\,.
\end{align}
For a hyper-cubic lattice in $D$ dimensions, the 
correlation function (\ref{corrbb}) can be expressed with
modified Bessel functions $I_\alpha(z)$,
\begin{align}
&\langle \hat{b}^\dagger_\mu(t)\hat{b}_\nu(t')\rangle_\mathrm{therm}
=-\frac{1}{\pi}\int d\omega\,\mathcal{P}\frac{1}{e^{\beta \omega}-1} \int_0^\infty ds e^{-s+i\omega(t-t')}\nonumber\\
&\times\mathrm{Im}\left[ G(\omega+i\epsilon) \prod_{j=1}^D 
(-1)^{\triangle \mathbf{x}_{\mu\nu}^j}I_{\triangle \mathbf{x}_{\mu\nu}^j}
\left(\frac{J G(\omega+i\epsilon)s}{D}\right)\right]\,,
\end{align}
where $\triangle \mathbf{x}_{\mu\nu}^j=x_{\mu}^j-x_{\nu}^j$
is the $j$-component of the distance between two lattice sites.
$\triangle \mathbf{x}_{\mu\nu}^j$ is always an integer since we rescaled the distances with the lattice constant.

\begin{widetext}

\begin{figure}
\includegraphics[width=17cm]{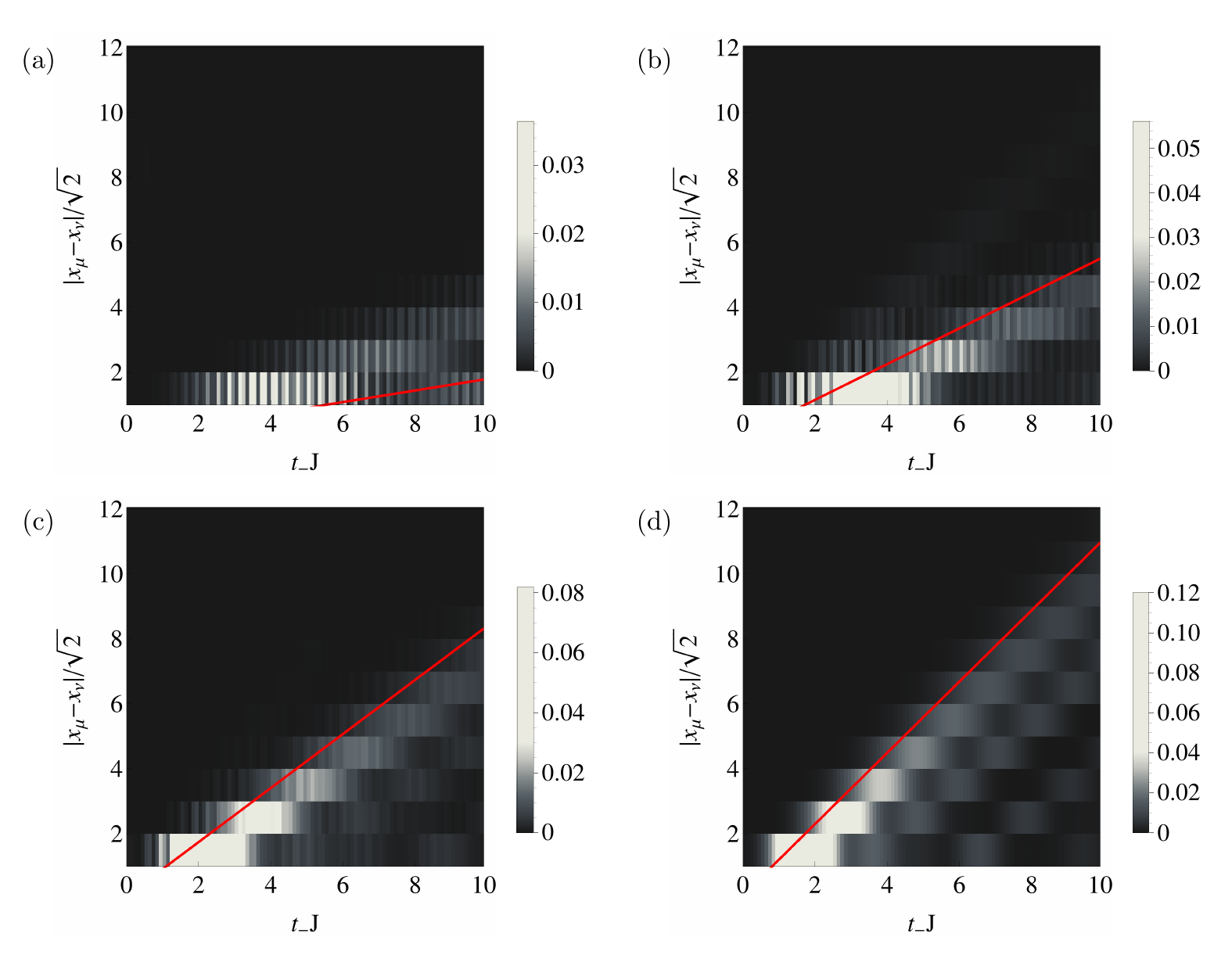}
\caption{Time-dependence of the two-time correlation functions $|\langle \hat{b}^\dagger(t)\hat{b}(t')\rangle_\mathrm{therm}|^2$ 
along the diagonal for a lattice in two dimensions for unit filling and $J/U=0.1$.
The parameters are (a) $\beta U=1$, (b) $\beta U=3$, (c) $\beta U=5$, (d) $\beta U=\infty$.
The maximum velocity of the correlations decreases with the temperature 
of the thermal ensemble.
When the temperature grows, also highly energetic modes contribute to the dynamics.
The amplitude of the correlations decreases with increasing temperature.
}\label{diagcorrdoubletime}
\end{figure}

\end{widetext}

In Fig.~\ref{diagcorrdoubletime} we evaluate the double-time 
correlation function for different values of $\beta U$.
The density plots show a temperature-dependent light-cone structure.
For low temperatures it is viable to neglect occupation numbers $n>2$.
The dynamics of the correlation functions w.r.t.
to the time difference $t_-=(t-t')/2$ at unit filling
is determined by the frequencies 
\begin{align}\label{partholetherm}
\Omega_\mathbf{k}=&U-J(1-3p_0) T_\mathbf{k}\nonumber\\
&-\sqrt{U^2-6 J(1-3p_0) U T_\mathbf{k}+J^2(1-3p_0)^2T_\mathbf{k}^2}
\end{align}
with $p_0=p_2=1/(2+e^{\beta U/2})$ being the probability to find zero or two particles on a lattice site.
The maximum group velocity, given by 
$v^\mathrm{max}=\mathrm{max}_\mathbf{k}|\nabla_\mathbf{k} \Omega_\mathbf{k}|$,
determines the light-cone structure of the correlations.
At zero temperature, the propagation velocity of the 
correlations is maximal, see Fig.~\ref{diagcorrdoubletime} (d).
In this limit, only the doublon-holon excitations $\langle P^{10}_\mu(t)P^{12}_\nu(t')\rangle$,  
$\langle P^{21}_\mu(t)P^{01}_\nu(t')\rangle$,  
$\langle P^{10}_\mu(t)P^{01}_\nu(t')\rangle$,
and
$\langle P^{21}_\mu(t)P^{12}_\nu(t')\rangle$
are relevant. 

For finite temperatures, we have $1-3p_0<1$.
According to equation (\ref{partholetherm}), this implies  
a shrinking propagation velocity, see Fig.~\ref{diagcorrdoubletime} (c).
If the temperature is increased even further, the 
approximation of thermal particle-hole excitations
is not valid anymore since also occupation numbers $n>2$ play a role.
Therefore, the analytical result (\ref{partholetherm}) 
does not determine the light-cone structure anymore, 
see Fig.~\ref{diagcorrdoubletime} (a)~and~(b).

\begin{figure}[h]
\includegraphics[width=6.8cm]{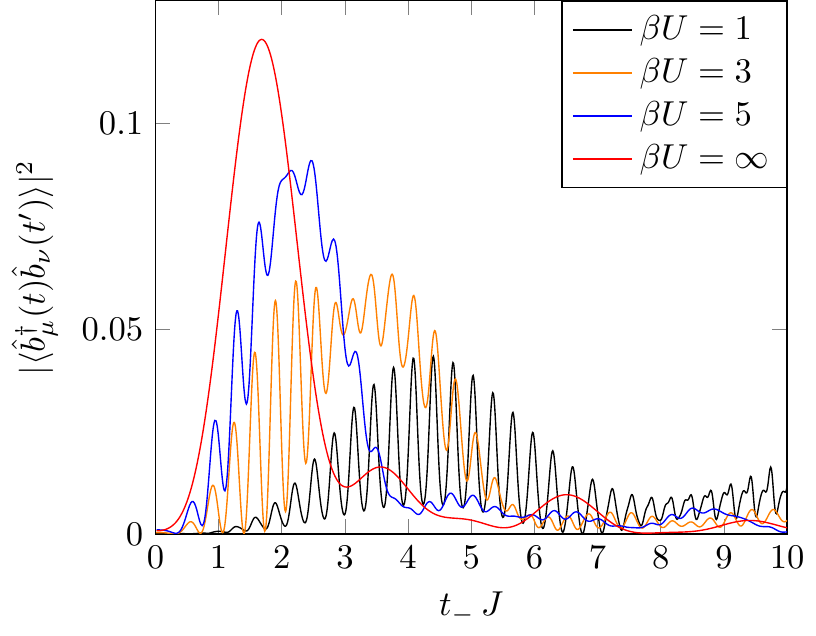}
\caption{Double-time correlation for $|x_\mu-x_\nu|=\sqrt{2}$ in two dimensions 
at different temperatures.
The position of the maximum is shifting with temperature 
approximately as $(t_-J)\sim 1/(1-3 p_0)$.
}\label{ThermalCorrl1}
\end{figure}

The population of higher excited states is also reflected 
in the time-evolution of the correlations at a fixed distance $|x_\mu-x_\nu|$.
At $\beta U=\infty$, only eigen-modes of order $\mathcal{O}(U)$ play a role whereas 
for large temperatures the eigen-modes of order $\mathcal{O}(nU)$,~$n>1$
dominate the time-evolution, see Fig.~\ref{ThermalCorrl1}.

 \newpage

\section{Equilibration after a quantum quench}\label{BHnonequil1}

In this section we turn to the non-equilibrium dynamics of excitations 
in the Bose-Hubbard system after a quantum quench at finite 
temperatures from $J=0$
to $J>0$.
The evolution of time-local quantities
can be deduced from double-time hierarchy for $t=t'$ 
of from the single-time hierarchy in \cite{NS10}.
Afterwards we compare the pre-thermalized expectation 
values at $t\rightarrow \infty$ with the corresponding thermal expectation values.

We assume the system to be initially in a 
thermal state which is determined by the 
density matrix
\begin{align}
\hat{\rho}^\mathrm{in}=\bigotimes_\mu \hat\rho_\mu, \quad
\hat\rho_\mu=\sum_n p_n|n\rangle_\mu \langle n|\,, 
\end{align}
where the $p_n$ are given by the thermal distribution (\ref{thermalpn}).
The hopping rate is switched suddenly from $J=0$
to a finite value which is still in the 
Mott regime $J<J_\mathrm{crit}$.
The dynamics of the on-site probabilities follow the equations
\begin{align}\label{onsite}
&i\partial_t \langle \hat{P}^{n,n}_\mu\rangle=-\frac{J}{Z}\sum_{\kappa}T_{\mu\kappa}
\sum_m \sqrt{m+1}\\
\times&\bigg[\sqrt{n}\big(\langle \hat{P}_\mu^{n,n-1}\hat{P}_\kappa^{m,m+1}\rangle^\mathrm{corr}
-\langle \hat{P}_\kappa^{m+1,m}\hat{P}_\mu^{n-1,n}\rangle^\mathrm{corr}\big)\nonumber\\
&-\sqrt{n+1}\big(\langle \hat{P}_\mu^{n+1,n}\hat{P}_\kappa^{m,m+1}\rangle^\mathrm{corr}
-\langle \hat{P}_\kappa^{m+1,m}\hat{P}_\mu^{n,n+1}\rangle^\mathrm{corr}\big)\bigg]\nonumber\,,
\end{align}
and the two-point correlations evolve according to
%
%
\begin{align}\label{twosites}
&[i\partial_t+U(n-m)]\langle \hat{P}^{n+1,n}_\mu\hat{P}^{m,m+1}_\nu\rangle^\mathrm{corr}\nonumber\\
=-&\frac{J}{Z}\sqrt{n+1}(\langle \hat{P}^{n+1,n+1}_\mu\rangle-\langle \hat{P}^{n,n}_\mu\rangle)\nonumber\\
&\times\sum_{\kappa\neq\mu,\nu}T_{\kappa\mu}
\langle \hat{b}^\dagger_\kappa\hat{P}^{m,m+1}_\nu\rangle^\mathrm{corr}\nonumber\\
-&\frac{J}{Z}\sqrt{m+1}(\langle \hat{P}^{m,m}_\nu\rangle-\langle \hat{P}^{m+1,m+1}_\nu\rangle)\nonumber\\
&\times\sum_{\kappa\neq\mu,\nu}T_{\kappa\nu}
\langle \hat{P}^{n+1,n}_\mu \hat{b}_\kappa\rangle^\mathrm{corr}\nonumber\\
-&\frac{J}{Z}T_{\mu\nu}\sqrt{n+1}\sqrt{m+1}\langle \hat{P}^{n+1,n+1}_\mu\rangle\langle \hat{P}^{m,m}_\nu\rangle\nonumber\\
+&\frac{J}{Z}T_{\mu\nu}\sqrt{n+1}\sqrt{m+1}\langle \hat{P}^{n,n}_\mu\rangle\langle \hat{P}^{m+1,m+1}_\nu\rangle\,.
\end{align}
In the limit of zero temperature, the doublon-holon correlation 
functions decouple from correlation functions which involve 
occupation numbers $n>2$.
This case has been discussed in references \cite{NS10,QKNS14}.
However, we shall see that a finite temperature alters
the pre-thermalization dynamics.

\begin{widetext}
 
\begin{figure}
\includegraphics[width=18cm]{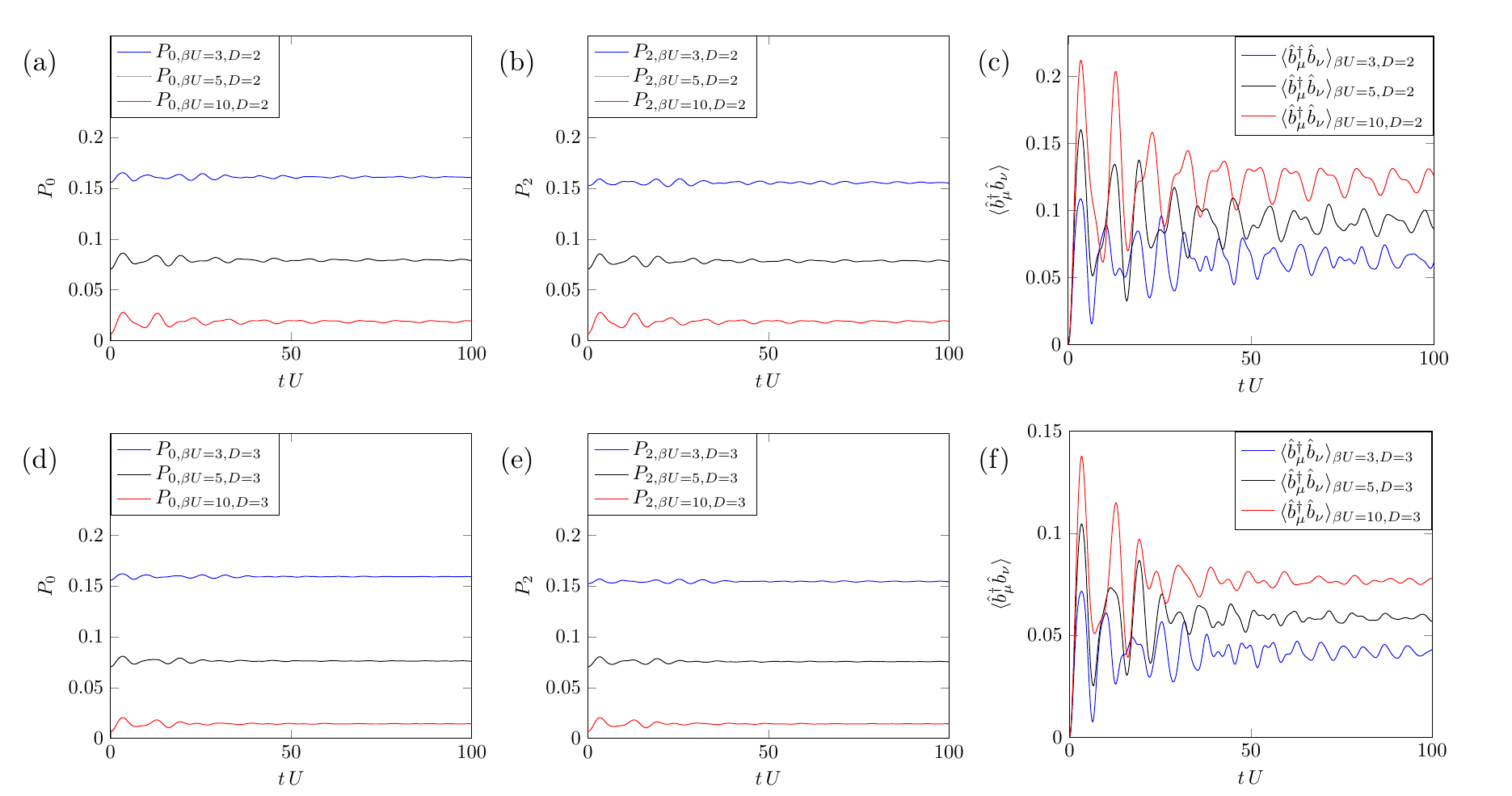}
\caption{Equilibration dynamics in two and three dimensions.
The amplitude of the oscillations decreases with increasing temperature 
of the initial state.
For all initial temperatues, the system prethermalizes faster in three dimensions than it does 
in two dimensions.
In our computation, we did not restrict the occupation number.
Nevertheless, we see that also for finite but sufficiently low temperatures, the 
particle-hole symmetry is approximately valid.
}\label{dynamics2D3D}
\end{figure}

\end{widetext}

The equations (\ref{onsite}) and (\ref{twosites})
are nonlinear due to the back-reaction of the 
on-site quantities onto the correlations.
This back-reaction can be neglected since 
it is of order $1/Z^2$ and we have 
$\langle \hat P^{n,n}_\mu\rangle=P_n=p_n+\mathcal{O}(1/Z)$.
Exploiting the translational symmetry, 
we can simplify these equations by a spatial 
Fourier transformation
\begin{align}\label{Fourier}
\langle \hat{P}_\mu^{n,n-1} \hat{P}_\nu^{m-1,m}\rangle=\frac{1}{N}\sum_\mathbf{k}f^{n,m}_\mathbf{k}e^{i\mathbf{k}\cdot(\mathbf{x}_\mu-\mathbf{x}_\nu)}\,,
\end{align}
where $N$ denotes the number of lattice sites.
Now, the on-site probabilities $P_n=\langle \hat{P}^{n,n}_\mu\rangle $ evolve according to
\begin{align}\label{pn}
i\partial_t  P_n=&-\frac{J}{N}\sum_{\mathbf{k}}T_\mathbf{k}
\sum_m \sqrt{m+1}\nonumber\\
&\times\bigg[\sqrt{n}\big(f_\mathbf{k}^{n,m+1}
-f_\mathbf{k}^{m+1,n}\big)\nonumber\\
&-\sqrt{n+1}\big(f_\mathbf{k}^{n+1,m+1}
-f_\mathbf{k}^{m+1,n+1}\big)\bigg]\,,
\end{align}
whereas for the Fourier components of the correlations 
we find
\begin{align}\label{bb}
&[i\partial_t+U(n-m)]f_\mathbf{k}^{n+1,m+1}\nonumber\\
&=-JT_\mathbf{k}\sqrt{n+1}(p_{n+1}-p_n)\sum_{s=0}^\infty 
\sqrt{s+1}f_\mathbf{k}^{s+1,m+1}\nonumber\\
&-JT_\mathbf{k}\sqrt{m+1}(p_{m}-p_{m+1})\sum_{s=0}^\infty 
\sqrt{s+1}f_\mathbf{k}^{n+1,s+1}\nonumber\\
&-JT_\mathbf{k}\sqrt{m+1}\sqrt{n+1}(p_{n+1}p_m-p_{n}p_{m+1})\,.
\end{align}
We solved the coupled system of equations 
in two and three dimensions for various temperatures, see Fig.~\ref{dynamics2D3D}.
The on-site probabilities are oscillating around their initial 
thermal value $p_n$.
Especially for high temperatures, the back action of these 
quantities onto the correlation functions can indeed be 
safely neglected since\linebreak $|P_n(t)-P_n(0)|/P_n(0)\ll1$.
In order to gain some analytical insight 
into the equilibration process, we
neglect $p_n$ for $n>2$.
As a consequence, we obtain the particle-hole 
symmetry $p_0=p_2$.
When the system is quenched from a thermal state, 
the oscillations decrease with 
time and the on-site probabilities 
approach for $t\rightarrow\infty $
\begin{align}\label{pnequil}
&P_{0/2,\mathrm{equil}}= p_{0}+\frac{1}{N}\sum_\mathbf{k}
\frac{4J^2T_\mathbf{k}^2(1-4p_0+3p_0^2)}{\omega_\mathbf{k}^2}
\end{align}
with
\begin{align}
\omega_\mathbf{k}=&
\sqrt{\big[U^2-6JT_\mathbf{k}(1-3p_0)U+J^2T_\mathbf{k}^2(1-3p_0)^2\big]}\,.
\end{align}
For the lattice-site correlations we for $t\rightarrow\infty $ the asymptotic 
expression
\begin{align}\label{correquil}
&\langle \hat{b}^\dagger_\mu\hat{b}_\nu\rangle_\mathrm{equil}=
\frac{1}{N}\sum_\mathbf{k}e^{i\mathbf{k}\cdot(\mathbf{x}_\mu-\mathbf{x}_\nu)}
\frac{4JT_\mathbf{k}U(1-4p_0+3p_0^2)}{\omega_\mathbf{k}^2}\,.
\end{align}
Given equations (\ref{pnequil}) and (\ref{correquil}), we conclude that the quench-induced 
change of these quantities becomes smaller with increasing 
temperature.

\begin{center}
\begin{figure}[h]
\includegraphics[width=7.3cm]{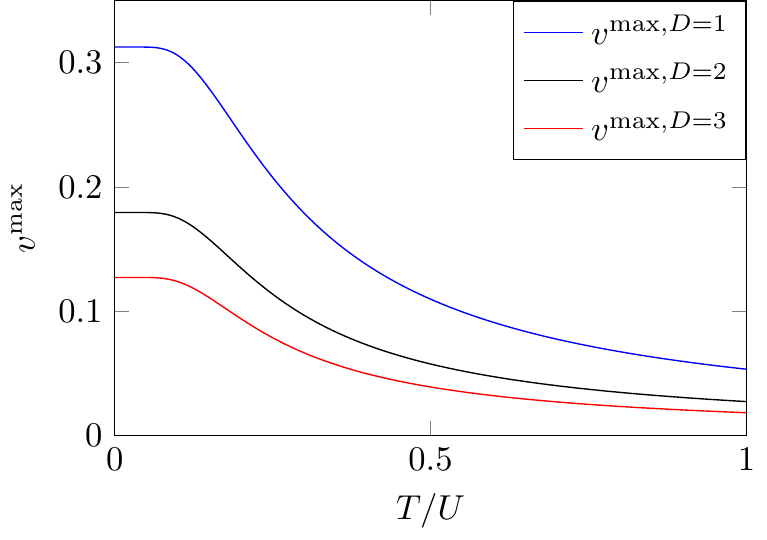}
\caption{The maximum velocity of the correlations after a quantum quench 
from $J=0$ to $J/U=0.1$ depends on the temperature of the initial 
ensemble.}\label{vmax}
\end{figure}
\end{center}

From the eigen-modes
we can also estimate the maximum propagation 
speed of the expanding correlations.
%
As before, we can estimate the maximum propagation 
velocity from $v^\mathrm{max}=\mathrm{max}_\mathbf{k}|\nabla_\mathbf{k}\Omega_\mathbf{k}|$.
%
In a hyper-cubic lattice in $D$ dimensions 
with small $J/U$ we have  
$v^\mathrm{max}=J(3-9p_0)/D$ along the lattice axes
and $v^\mathrm{max}=J(3-9p_0)/\sqrt{D}$ along the diagonals.
As before, the maximum propagation speed becomes smaller 
for increasing temperatures, see Fig.~\ref{vmax}.

The pre-thermalized state is a quasi-stationary state which is not 
necessarily thermal whereas real thermalization occurs on much longer 
time-scales \cite{BBS03,BBW04,KWE11}.
However, we can compare the pre-thermalized state 
with a corresponding thermal state that has the 
same temperature $T$ and filling $\langle \hat{n}_\mu\rangle $.
In order to do so, we compute the asymptotic values 
of the on-site probabilities $P_{n,\mathrm{equil}} $ and the correlation functions 
$\langle \hat{b}_\mu^\dagger\hat{b}_\nu\rangle_\mathrm{equil}$
from equations (\ref{pn}) and (\ref{bb}) without further approximations.
The thermal correlation $\langle \hat{b}_\mu^\dagger\hat{b}_\nu\rangle_\mathrm{therm}$ 
can be deduced from equation (\ref{corrbb}).
For the order $\mathcal{O}(1/Z)$-contribution of the $P_{n,\mathrm{therm}} $ we take
the expression from reference \cite{NQS16}.
In this article, we evaluated the partition function 
of the Bose-Hubbard system up to $\mathcal{O}(1/Z)$ and derived 
various thermal quantities.
Explicitly we have (see equation 21 in \cite{NQS16})
%
\begin{align}
P_{n,\mathrm{therm}}=&p_n+\frac{1}{N}\sum_\mathbf{k}\sum_{l=-\infty}^{\infty}
\frac{J T_\mathbf{k}}{1+JT_\mathbf{k}G(\omega_l)}\bigg[p_n G(\omega_l)\nonumber\\
&-\frac{(n+1)p_n}{\omega_l -U n+\mu}+\frac{np_n}{\omega_l-U(n-1)+\mu}\nonumber\\
&+\frac{n(p_{n-1}-p_n)}{\beta(\omega_l-U(n-1)+\mu)^2)^2}\nonumber\\
&-\frac{(n+1)(p_n-p_{n+1})}{\beta(\omega_l-U n+\mu)^2}\bigg]
\end{align}
with the bosonic Matsubara modes $\omega_l=2\pi i l/\beta$.

\begin{widetext}

\begin{figure}
\includegraphics[width=17cm]{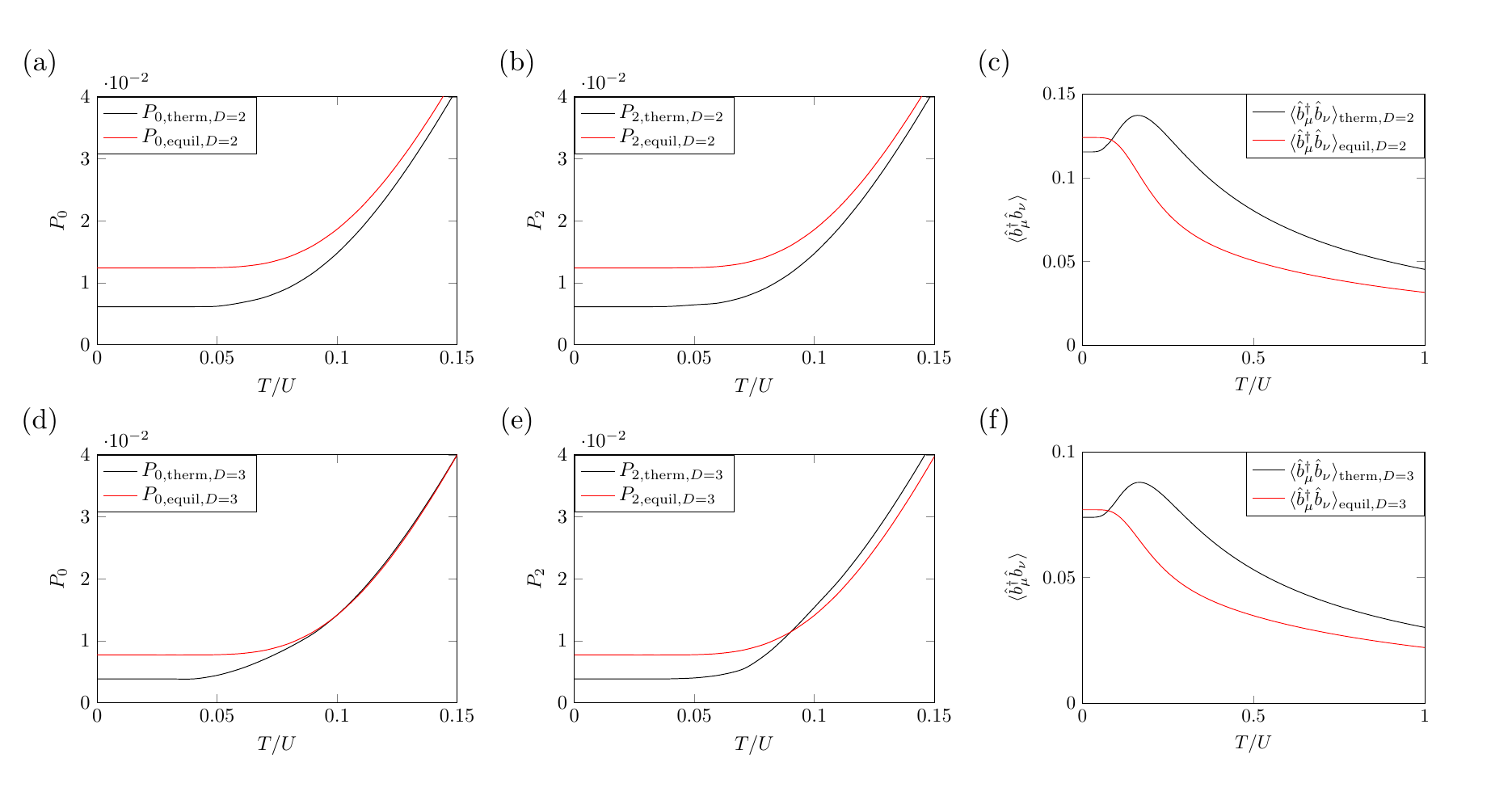}
\caption{After a quantum quench at a temperature $T $, the occupation probabilities and the correlation functions approach a prethermalized value (red curves).
This is compared with the corresponding thermal expressions (black curves).
For low temperatures, the onsite-quantities differ roughly by a factor of 2. 
The difference between prethermalized and thermal 
values diminishes with increasing temperatures.
Note that deviations from the particle-hole symmetry at high temperatures are larger
in three dimensions (see (d) and (e)) than in two dimensions (see (d) and (e)).
In (c) and (f) we depict the temperature-dependence of the prethermalized and thermal expectation values of the next neighbour correlations $\langle \hat{b}_\mu^\dagger\hat{b}_\nu\rangle$.
Note that the thermal correlation function has its global maximum at $T>0$.
}\label{thermalasym2D3D}
 \end{figure}

\end{widetext}

%
%
At zero temperature, the values 
differ roughly by a factor of 2 which has been 
stated elsewhere \cite{MK09}.
For sufficiently low temperatures, 
the particle-hole symmetry is valid, $P_0\sim P_2$, which 
can be deduced from comparing Fig.~\ref{thermalasym2D3D} (a)
and (b) or Fig.~\ref{thermalasym2D3D} (d) and (e).
The energy which is transferred 
by a quench to the system becomes 
more and more irrelevant for increasing 
temperatures.
Thus, the difference between thermal and pre-thermalized 
quantities reduces with growing $T$.
From Fig.~\ref{thermalasym2D3D} (c) and (f) we see 
how the increasing temperature diminishes the correlations
between lattice sites. 
However, the thermal next neighbor correlation 
function has its global maximum at finite $T>0$.

 \newpage

\section{Quantum quench and Double-time correlations}\label{BHnonequil2}

A quantum quench changes also the light-cone structure of the double-time correlation functions.
We assume that before the quantum quench the system is in an equilibrium 
state with a finite hopping rate $J_-$.
At time $t=0$, the hopping rate is quenched to a value 
$J_+<J_\mathrm{crit}$.
For simplicity, we restrict our considerations 
to zero temperature.
Analogous to equation (\ref{Fourier}) we define the 
Fourier transform of the two-time correlations
\begin{align}
\langle \hat{P}_\mu^{n,n-1}(t) \hat{P}_\nu^{m-1,m}(t')\rangle=\frac{1}{N}\sum_\mathbf{k}f^{n,m}_\mathbf{k}(t,t')e^{i\mathbf{k}\cdot(\mathbf{x}_\mu-\mathbf{x}_\nu)}\,.
\end{align}
The dynamics of the correlation functions follows directly from equation (\ref{twosite})
and a corresponding relation which takes into account the time-evolution w.r.t.~variable $t'$.
Neglecting the back-reaction onto the on-site probabilities, the Fourier coefficients
evolve up to order $1/Z$ according to
\begin{align}\label{odet1}
i\partial_t f^{11}_\mathbf{k}&= -J^\pm T_\mathbf{k}(f_\mathbf{k}^{11}+\sqrt{2}f_\mathbf{k}^{21})-J^\pm T_\mathbf{k}\,,\\
i\partial_t f^{12}_\mathbf{k}&= -J^\pm T_\mathbf{k}(f_\mathbf{k}^{12}+\sqrt{2}f_\mathbf{k}^{22})\,,\\
i\partial_t f^{21}_\mathbf{k}&= \sqrt{2}J^\pm T_\mathbf{k}(f_\mathbf{k}^{11}+\sqrt{2}f_\mathbf{k}^{21})-U f_\mathbf{k}^{21}+\sqrt{2}J^\pm T_\mathbf{k}\,,\\
i\partial_t f^{22}_\mathbf{k}&= \sqrt{2}J^\pm T_\mathbf{k}(f_\mathbf{k}^{12}+\sqrt{2}f_\mathbf{k}^{22})-U f_\mathbf{k}^{22}\,,
\end{align}
and
\begin{align}\label{odet2}
i\partial_{t'} f^{11}_\mathbf{k}&= J^\pm T_\mathbf{k}(f_\mathbf{k}^{11}+\sqrt{2}f_\mathbf{k}^{12})+J^\pm T_\mathbf{k}\,,\\
i\partial_{t'} f^{12}_\mathbf{k}&= -\sqrt{2}J^\pm T_\mathbf{k}(f_\mathbf{k}^{11}+\sqrt{2}f_\mathbf{k}^{12})+U f_\mathbf{k}^{12}-\sqrt{2}J^\pm T_\mathbf{k}\,,\\
i\partial_{t'} f^{21}_\mathbf{k}&= J^\pm T_\mathbf{k}(f_\mathbf{k}^{21}+\sqrt{2}f_\mathbf{k}^{22})\,,\\
i\partial_{t'} f^{22}_\mathbf{k}&= -\sqrt{2}J^\pm T_\mathbf{k}(f_\mathbf{k}^{21}+\sqrt{2}f_\mathbf{k}^{22})+U f_\mathbf{k}^{22}
\,.
\end{align}
For $t,t'<0$, the correlation functions can be immediately obtained from the ground state 
correlations at $J=J_-$ \cite{QKNS14}.
Due to the quantum quench, the correlation function is not anymore homogeneous in 
time but depends on the relative time $t_-=(t-t')/2$
and on the central time $t_+=(t+t')/2$.
For $t_+ \pm t_-<0$ and $t_+ \pm t_->0$ we find correlations before and after the hopping quench, respectively. 
Correlations between an creation (annihilation) of a 
particle before the quench and an annihilation (creation)
of particle can be obtained for $t_+ + t_-<0$ and 
$t_+ - t_->0$ ($t_+ + t_->0$ and 
$t_+ - t_-<0$).

\begin{figure}
\includegraphics[width=7cm]{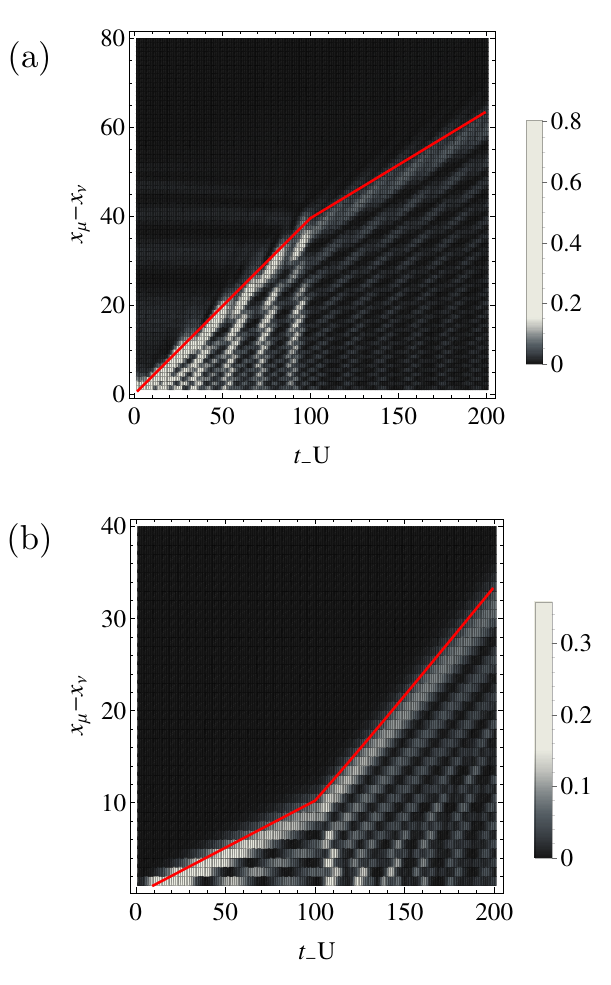}
\caption{Double-time correlation function $\langle \hat{b}^\dagger_\mu (t)\hat{b}_\nu (t')\rangle_{T=0}$
in one dimension for a quench at $t=0$ from $J_-/U=0.05$ to $J_+/U=0.15$.
(a): $t_+U=100$, the lightcone has a kink at $t_-=t_+$ where the maximum velocity 
changes from $v^\mathrm{max}_{++}\approx0.40$ to $v^\mathrm{max}_{+-}\approx0.24$.
(b): $t_+U=-100$, the lightcone has a kink at $t_-=-t_+$ where the maximum velocity 
changes from $v^\mathrm{max}_{--}\approx 0.10$ to $v^\mathrm{max}_{+-}\approx0.24$.
}\label{doubletimecorr}
\end{figure}

The light-cone structure is primarily determined by three different velocities.
For $t,t'<0$, the maximum velocity $v^\mathrm{max}_{--}=\mathrm{max}_\mathbf{k}|\nabla_\mathbf{k} \Omega^{--}_\mathbf{k}|$ can be estimated from 
\begin{align}
\Omega_\mathbf{k}^{--}=U-J_-T_\mathbf{k}-\sqrt{U^2-6 J_- UT_\mathbf{k}+(J_-T_\mathbf{k})^2}\,. 
\end{align}
For $t>0$ and $t'<0$ or $t<0 $ and $ t'>0$, the spread is determined by
both hopping rates $J_-$ and $J_+$.
Thus we have the eigen-modes
\begin{align}
\Omega_\mathbf{k}^{+-}=&\frac{1}{2}\left(U-J_-T_\mathbf{k}-\sqrt{U^2-6 J_- UT_\mathbf{k}+(J_-T_\mathbf{k})^2}\right)\nonumber\\
+&\frac{1}{2}\left(U-J_+T_\mathbf{k}-\sqrt{U^2-6 J_+ UT_\mathbf{k}+(J_+T_\mathbf{k})^2}\right)\,.
\end{align}
from which one can obtain $v^\mathrm{max}_{+-}=\mathrm{max}_\mathbf{k}|\nabla_\mathbf{k} \Omega^{+-}_\mathbf{k}|$.
Finally, if $t,t'>0$, the spread of correlations is dominantly determined by $J_+$.
Here the maximum velocity $v^\mathrm{max}_{++}=\mathrm{max}_\mathbf{k}|\nabla_\mathbf{k} \Omega^{++}_\mathbf{k}|$
can be derived from
\begin{align}
\Omega_\mathbf{k}^{++}=U-J_+T_\mathbf{k}-\sqrt{U^2-6 J_+ UT_\mathbf{k}+(J_+T_\mathbf{k})^2}\,. 
\end{align}
The kink in the light-cone structure due to the quantum quench 
is illustrated in Fig.~\ref{doubletimecorr}.
The complete light-cone structure for the double-time correlators 
in arbitrary dimensions is evaluated in appendix \ref{doubtime}.

\section{Conclusions}

We studied equilibrium properties and non-equilibrium 
dynamics of the Bose-Hubbard model in the Mott insulating phase.
To this end, we extended the hierarchy for large coordination 
numbers $Z$ presented in \cite{NS10} to a hierarchy for double-time 
correlation functions.

This enabled us to derive thermal correlation functions 
and we studied the spread of two-time correlation functions at 
finite initial temperatures, see Figs.~\ref{diagcorrdoubletime} and \ref{ThermalCorrl1}.
We restricted our considerations to first order in $1/Z$ which 
governs the free quasi-particle evolution.
As usual, the effective light-cone structure was obtained via a
saddle-point approximation.

As we demonstrated above, the dynamics of strongly interacting 
quantum many-body systems and their thermal properties can be 
both accessed within our approach.
The phenomenon of pre-thermalization after a quantum quench 
was generalized from $T=0$ \cite{QKNS14} to finite temperatures, 
see Figs.~\ref{dynamics2D3D} and \ref{vmax}.
Note that the dynamical equations in order $\mathcal{O}(1/Z)$ 
cover only the short-time evolution of the excitations.
In order to include the long-time evolution which 
is determined by quasi-particle scattering, 
also higher order correlations have to be taken into account.
For real thermalization processes, for example, one has to 
evaluate the hierarichal equations up to order 
$\mathcal{O}(1/Z^3)$ in order to obtain a Boltzmann equation 
\cite{QS19a}.

Apart from these theoretical investigations, our approach can 
also be applied in order to interpret experimental results.  
Since the double-time hierarchy induces a Green function 
hierarchy, the latter can used for the interpretation of 
pump-probe experiments \cite{}.
An excitation which is created by a pump beam at time $t$ 
and measured by a probe beam at time $t'$ is naturally 
described in terms of the lesser Green functions \cite{ATE14,FMSSD17,FKP17}.
Moreover, the Green function hierarchy allows for a controlled
perturbative calculation of the self-energy or the 
susceptibility of strongly correlated systems.

As an outlook, the study of out-of-time-correlators (OTOC), 
a common measure for quantum chaos, should be feasible within our approach \cite{LO69,MSS16}.
In this context one has to extend the double-time hierarchy of correlations to
a hierarchy of multi-time correlations.
This should be possible in complete analogy to the approach 
presented in this work and will be the subject of further studies.

\acknowledgments


This work was funded by DFG, grant \# 278162697 (SFB 1242) and 398912239.

 \begin{widetext}

\section{Appendix: Proof of the hierarchy}\label{PH}

The most general double-time expectation value has the form $\langle 
\hat{A}_\mathcal{S}(t)\hat{B}_\mathcal{S}(t')\rangle$ where 
$\mathcal{S}$ denotes the set of $|\mathcal{S}|$ lattice sites.
In order to avoid cluttering of indices, we introduce a short hand notation for 
operators which are defined on the set $\mathcal{S}$ via 
$\hat{A}_\mathcal{S}(t)=\hat{A}^1_{\mu_1}(t)\times...\times A^n_{\mu_n}(t)$
and $\hat{B}_\mathcal{S}(t')=\hat{B}^1_{\mu_1}(t')\times...\times \hat{B}^n_{\mu_n}(t')$.
From the equation-of-motion hierarchy for the expectation values 
$\langle \hat{A}_\mathcal{S}(t)\hat{B}_\mathcal{S}(t')\rangle$,
the hierarchy for expectation values of the form 
$\langle \hat{A}(t)_{\mathcal{P}_1} \hat{B}_{\mathcal{P}_2}(t')\rangle$ 
can be obtained by setting $\mathcal{S}=\mathcal{P}_1 \cup\mathcal{P}_2$
and choosing $\hat{A}_\mu(t)=\hat{1}_\mu$ for $\mu\notin \mathcal{P}_1$ and 
$\hat{B}_\nu(t')=\hat{1}_\nu$ for $\nu\notin \mathcal{P}_2$.

In \cite{NS10}, we presented a hierarchy for the correlated 
parts of a many-particle density matrix based on a large coordination number $Z$.
This hierarchy was formulated for operators
but it can also be stated in terms of 
correlation functions.
In the following, we will derive the corresponding hierarchy 
for two-time correlation functions .
The starting point is a separation of the correlation functions 
in correlated and uncorrelated parts.
For two sites, the correlated part of an expectation values is defined by
as
\begin{align}
 \langle\hat{A}_\mu(t)\hat{B}_\nu(t)\hat{C}_\mu(t')\hat{D}_\nu(t')\rangle^\mathrm{c}
=\langle\hat{A}_\mu(t)\hat{B}_\nu(t)\hat{C}_\mu(t')\hat{D}_\nu(t')\rangle
-
\langle\hat{A}_\mu(t)\hat{C}_\mu(t')\rangle\langle\hat{B}_\nu(t) \hat{D}_\nu(t')\rangle\,.
\end{align}
%
Arbitrary correlation functions can be separated according to
\begin{align}\label{partitions}
\langle \hat{A}_{\cal S}(t)\hat{B}_{\cal S}(t')\rangle^\mathrm{c}&=\langle 
\hat{A}_{\cal S}(t)\hat{B}_{\cal S}(t')\rangle - 
\sum_{\cup_i {\cal{P}}_i={\cal{S}},{{\cal P}}_i\subset {\cal{S}}} \prod_{{\cal 
P}_i}\langle \hat{A}_{{\cal{P}}_i}(t)\hat{B}_{{\cal{P}}_i}(t')\rangle^\mathrm{c}\,,
\end{align}
where the sum is over all proper partitions of the set $\cal{S}$.
Note that the definition of the correlated parts agrees 
with the separation of the density matrix into correlated parts 
given in \cite{NS10} if the operators depending on $t$ (or $t'$)
are replaced by the unity operator.
The central point of our derivation is the scaling hierarchy of the 
correlations,
\begin{eqnarray}
\langle 
\hat{A}_\mathcal{S}(t)\hat{B}_\mathcal{S}(t')\rangle^\mathrm{c}=\mathcal{O}(Z^{
1-|\mathcal{S}|})\,.
\end{eqnarray}
Rewriting the hierarchy of the density matrix from reference \cite{NS10}
in terms of expectation values
it is possible to guess the corresponding hierarchy for 
for the equations of motion of 
$\langle\hat{A}_\mathcal{S}(t)\hat{B}_\mathcal{S}(t')\rangle^\mathrm{c}$.
We claim that the full hierarchy for an arbitrary set of $\mathcal{S}$ lattice sites 
has the form

\begin{subequations}
\label{equations}
\begin{align}
\label{hierarchy1}
i \partial_t\langle \hat{A}_{\cal S}(t)\hat{B}_{\cal S}(t')\rangle^\mathrm{c}
&=\sum_{\mu \in {\cal S}} \langle 
[\hat{A}_{\mathcal{S}}(t),\hat{H}_\mu(t)]\hat{B}_{\cal S}(t')\rangle^\mathrm{c}
-\frac{J}{Z}\sum_{\mu,\nu\in\mathcal{S}}T_{\mu\nu}\langle 
[\hat{A}_{\mathcal{S}}(t),\hat{X}_\mu^\dagger(t)\hat{X}_\nu(t)]\hat{B}_{\cal 
S}(t')\rangle^\mathrm{c}  \\
\label{hierarchy2}
&\hspace{-2cm}-\frac{J}{Z}\sum_{\kappa\notin\mathcal{S}}\sum_{\mu\in 
S}T_{\mu\kappa}
\bigg\{\langle 
\hat{X}^\dagger_\kappa(t)[\hat{A}_{\mathcal{S}}(t),\hat{X}_\mu(t)]\hat{B}_{
\mathcal{S}}(t')\rangle^\mathrm{c}+
\langle 
\hat{X}_\kappa(t)[\hat{A}_{\mathcal{S}}(t),\hat{X}_\mu^\dagger(t)]\hat{B}_{
\mathcal{S}}(t')\rangle^\mathrm{c}
\bigg\}\\
\label{hierarchy3}
&\hspace{-2cm}-\frac{J}{Z}\sum_{\kappa\notin\mathcal{S}}\sum_{\mu\in 
\mathcal{S}}\sum_{\mathcal{P}\subseteq\mathcal{S}\setminus\{\mu\}}^
{\mathcal{P}\cup\bar{\mathcal{P}}=\mathcal{S}\setminus\{\mu\}}T_{\mu\kappa}
\bigg\{
\langle[\hat{A}_{\mathcal{P}\cup\{\mu\}}(t),\hat{X}_\mu(t)]
\hat{B}_{\mathcal{P}\cup\{\mu\}}(t')\rangle^\mathrm{c}\langle\hat{X}
^\dagger_\kappa(t)\hat{A}_{\bar{\mathcal{P}}}(t)\hat{B}_{\bar{\mathcal{P}}}(t')
\rangle^\mathrm{c}\\
\label{hierarchy4}
&\hspace{1.5cm}+
\langle[\hat{A}_{\mathcal{P}\cup\{\mu\}}(t),\hat{X}_\mu^\dagger(t)]
\hat{B}_{\mathcal{P}\cup\{\mu\}}(t')\rangle^\mathrm{c}\langle\hat{X}_\kappa(t)\hat{A
}_{\bar{\mathcal{P}}}(t)\hat{B}_{\bar{\mathcal{P}}}(t')\rangle^\mathrm{c}
\bigg\}\\
\label{hierarchy5}
&\hspace{-2cm}-\frac{J}{Z}\sum_{\mu,\nu\in\mathcal{S}}\sum_{\mathcal{P}
\subseteq\mathcal{S}\setminus\{\mu,\nu\}}^{\mathcal{P}\cup\bar{\mathcal{P}}
=\mathcal{S}\setminus
\{\mu,\nu\}}T_{\mu\nu}\Bigg\{\langle 
[\hat{A}_{\mathcal{P}\cup\{\mu\}}(t),\hat{X}^\dagger_\mu(t)]\hat{B}_{\mathcal{P}
\cup\{\mu\}}(t')\rangle^\mathrm{c}
\langle 
\hat{X}_\nu(t)\hat{A}_{\bar{\mathcal{P}}\cup\{\nu\}}(t)\hat{B}_{\bar{\mathcal{P}
}\cup\{\nu\}}(t')\rangle^\mathrm{c}\\
\label{hierarchy6}
&\hspace{1.5cm}+\langle 
\hat{A}_{\mathcal{P}\cup\{\mu\}}(t)\hat{X}_\mu^\dagger(t)\hat{B}_{\mathcal{P}
\cup\{\mu\}}(t')\rangle^\mathrm{c}
\langle[\hat{A}_{\bar{\mathcal{P}}\cup\{\nu\}}(t),\hat{X}_\nu(t)]\hat{B}_{\bar{
\mathcal{P}}\cup\{\nu\}}(t')\rangle^\mathrm{c}\\
\label{hierarchy7}
&\hspace{-2cm}-\langle\hat{A}_{\mathcal{P}\cup\nu}(t)\hat{B}_{\mathcal{P}\cup\nu
}(t')\rangle^\mathrm{c}\Bigg[
\langle\hat{X}^\dagger_\nu(t)[\hat{A}_{\bar{\mathcal{P}}\cup\{\mu\}}(t),\hat{X}
_\mu(t)]
\hat{B}_{\bar{\mathcal{P}}\cup\{\mu\}}(t')\rangle^\mathrm{c}+
\langle\hat{X}_\nu(t)[\hat{A}_{\bar{\mathcal{P}}\cup\{\mu\}}(t),
\hat{X}^\dagger_\mu(t)]
\hat{B}_{\bar{\mathcal{P}}\cup\{\mu\}}(t')\rangle^\mathrm{c}\\
\label{hierarchy8}
&\hspace{-2cm}+\sum_{\mathcal{Q}\subseteq\bar{\mathcal{P}}}^{\mathcal{Q}\cup\bar
{\mathcal{Q}}=\bar{\mathcal{P}}}\bigg(
\langle[\hat{A}_{\mathcal{Q}\cup\{\mu\}}(t),\hat{X}_\mu(t)]
\hat{B}_{\mathcal{Q}\cup\{\mu\}}(t')\rangle^\mathrm{c}\langle 
\hat{X}^\dagger_\nu(t)\hat{A}_{\bar{\mathcal{Q}}}(t)\hat{B}_{\bar{\mathcal{Q}}}(t')
\rangle^\mathrm{c}\\
\label{hierarchy9}
&\hspace{1.5cm}+\langle[\hat{A}_{\mathcal{Q}\cup\{\mu\}}(t),\hat{X}
_\mu^\dagger(t)]
\hat{B}_{\mathcal{Q}\cup\{\mu\}}(t')\rangle^\mathrm{c}\langle 
\hat{X}_\nu(t)\hat{A}_{\bar{\mathcal{Q}}}(t)\hat{B}_{\bar{\mathcal{Q}}}(t')
\rangle^\mathrm{c}\bigg)\Bigg]
\Bigg\}
\end{align}
\end{subequations}
We will restrict ourselves to bosonic systems for proving
(\ref{hierarchy1})-(\ref{hierarchy9}) by induction.
For fermionic operators, additional signs will appear due to the permutations of the 
operators $\hat{A}_\mu$ and $\hat{B}_\mu$ in the partitions of the 
expectation values in correlated parts.

To begin, we assume that the hierarchy holds for all sets of lattice sites with cardinality \textit{strictly less} then $|\cal{S}|$.
From this we will derive the equations of motion for a set with cardinality \textit{equal} to $|\cal{S}|$.
The Heisenberg equations of motion for the operator $\hat{A}_{\cal{S}}$ lead to
\begin{eqnarray}\label{heisenberg}
i\partial_t \langle \hat{A}_{\cal{S}}(t)\hat{B}_{\cal{S}}(t')\rangle&=&
\sum_{\mu \in {\cal S}} \langle [\hat{A}_{\mathcal{S}}(t),\hat{H}_\mu(t)]\hat{B}_{\cal S}(t')\rangle-\frac{J}{Z}\sum_{\mu,\nu\in{\cal{S}}}T_{\mu\nu}
\langle[\hat{A}_{\cal{S}}(t),\hat{X}_\mu^\dagger(t)\hat{X}_\nu(t)]\hat{B}_{\cal{S}}(t')\rangle\nonumber\\
& &-\frac{J}{Z}\sum_{\kappa\notin \cal{S}}\sum_{\mu\in \cal{S}}T_{\mu\kappa}
\langle \hat{X}_\kappa^\dagger(t)[\hat{A}_{\cal{S}}(t),\hat{X}_\mu(t)]\hat{B}_{\cal{S}}(t')\rangle-\frac{J}{Z}\sum_{\kappa\notin \cal{S}}
\sum_{\mu\in \cal{S}}T_{\mu\kappa}\langle \hat{X}_\kappa(t)[\hat{A}_{\cal{S}}(t),\hat{X}_\mu^\dagger(t)]\hat{B}_{\cal{S}}(t')\rangle
\end{eqnarray}
As a next step, we separate the expectation values into correlated parts according to (\ref{partitions}).
The first term can be written as
\begin{eqnarray}\label{zero}
(0)\equiv\sum_{\mu \in {\cal S}} \langle [\hat{A}_{\mathcal{S}}(t),\hat{H}_\mu(t)]\hat{B}_{\cal S}(t')\rangle
=\sum_{{\cal P}\subseteq {\cal S}\setminus\{\mu\}}\langle [\hat{A}_{\mathcal{P}\cup\{\mu\}}(t),\hat{H}_\mu(t)]\hat{B}_{\mathcal{P}\cup\{\mu\}}(t')\rangle
\sum_{\cup_i{\cal{P}}_i=\cal{S}\setminus\cal{P}\cup\{\mu\}}
\prod_{{\cal{P}}_i}\langle \hat{A}_{{\cal{P}}_i}(t)\hat{B}_{{\cal{P}}_i}(t')\rangle^\mathrm{c}\,.
\end{eqnarray}
The second term on the right hand side of equation (\ref{heisenberg}) can be expanded as
\begin{eqnarray}\label{first}
& &-\frac{J}{Z}\sum_{\mu,\nu\in{\cal{S}}}T_{\mu\nu}
\langle[\hat{A}_{\cal{S}}(t),\hat{X}_\mu^\dagger(t)\hat{X}_\nu(t)]\hat{B}_{\cal{S}}(t')\rangle=(1)+(2a)+(2b)\nonumber\\
&=&-\frac{J}{Z}\sum_{\mu,\nu\in\cal{S}}T_{\mu\nu}\sum_{ {\cal{P}}\subseteq \cal{S}\setminus\{\mu,\nu\}}\Bigg\{\langle [\hat{A}_{\cal{P}\cup\{\mu\nu\}}(t),
\hat{X}_\mu(t)^\dagger \hat{X}_\nu(t)]\hat{B}_{\cal{P}\cup\{\mu\nu\}}(t')\rangle^\mathrm{c} \nonumber\\
& &+\sum_{\cal{Q}\subseteq{\cal{P}}}^{\cal{Q}\cup\bar{\cal{Q}}={\cal{P}}}\Bigg[
\langle[\hat{A}_{\cal{Q}\cup\{\mu\}}(t),\hat{X}^\dagger_\mu(t)]\hat{B}_{\cal{Q}\cup\{\mu\}}(t')\rangle^\mathrm{c} 
\langle \hat{X}_\nu(t)\hat{A}_{\bar{\cal{Q}}(t)\cup\{\nu\}}(t)\hat{B}_{\bar{\cal{Q}}\cup\{\nu\}}(t')\rangle^\mathrm{c}
\nonumber\\
& &+\langle \hat{A}_{\cal{Q}\cup\{\mu\}}(t)\hat{X}^\dagger_\mu(t)\hat{B}_{\cal{Q}\cup\{\mu\}}(t')\rangle^\mathrm{c}
\langle [\hat{A}_{\bar{\cal{Q}}\cup\{\nu\}}(t),\hat{X}_\nu(t)]\hat{B}_{\bar{\cal{Q}}\cup\{\nu\}}(t')\rangle^\mathrm{c}\Bigg]\Bigg\}
\sum_{\cup_i{\cal{P}}_i=\cal{S}\setminus\cal{P}\cup\{\mu,\nu\}}
\prod_{{\cal{P}}_i}\langle \hat{A}_{{\cal{P}}_i}(t)\hat{B}_{{\cal{P}}_i}(t')\rangle^\mathrm{c}
\end{eqnarray}
For brevety we denoted the three terms in equation (\ref{first}) with $(1)$, $(2a)$ and $(2b)$.
The third term in equation (\ref{heisenberg}) can be written as
\begin{eqnarray}\label{second}
& &-\frac{J}{Z}\sum_{\kappa\notin \cal{S}}\sum_{\mu\in \cal{S}}T_{\mu\kappa}
\langle \hat{X}_\kappa^\dagger(t)[\hat{A}_{\cal{S}}(t),\hat{X}_\mu(t)]\hat{B}_{\cal{S}}(t')\rangle=(3a)+(4a)\nonumber\\
&= &-\frac{J}{Z}\sum_{\kappa\notin \cal{S}}\sum_{\mu\in \cal{S}}T_{\mu\kappa}
\sum_{\cal{P}\subseteq \cal{S}\setminus\{\mu\}}\Bigg\{\langle \hat{X}_\kappa^\dagger(t)[\hat{A}_{\cal{P}\cup\{\mu\}}(t),\hat{X}_\mu(t)]\hat{B}_{\cal{P}\cup\{\mu\}}(t')\rangle^\mathrm{c}\nonumber\\
& &
+\sum_{\cal{Q}\subseteq \cal{P}}^{\cal{Q}\cup \bar{\cal{Q}}=\cal{P}}
\langle [\hat{A}_{\cal{Q}\cup\{\mu\}}(t),\hat{X}_\mu(t)]\hat{B}_{\cal{Q}\cup\{\mu\}}(t')\rangle^\mathrm{c} 
\langle \hat{X}_\kappa^\dagger(t)\hat{A}_{\bar{\cal{Q}}}(t)\hat{B}_{\bar{\cal{Q}}}(t')\rangle^\mathrm{c}\Bigg\}
\sum_{\cup_i{\cal{P}}_i=\cal{S}\setminus\cal{P}\cup\{\mu\}}
\prod_{{\cal{P}}_i}\langle \hat{A}_{{\cal{P}}_i}(t)\hat{B}_{{\cal{P}}_i}(t')\rangle^\mathrm{c}
\end{eqnarray}
Similarly, for the last summand in equation (\ref{heisenberg}) we have
\begin{eqnarray}\label{third}
& &-\frac{J}{Z}\sum_{\kappa\notin \cal{S}}\sum_{\mu\in \cal{S}}T_{\mu\kappa}
\langle \hat{X}_\kappa(t)[\hat{A}_{\cal{S}}(t),\hat{X}^\dagger_\mu(t)]\hat{B}_{\cal{S}}(t')\rangle=(3b)+(4b)\nonumber\\
&= &-\frac{J}{Z}\sum_{\kappa\notin \cal{S}}\sum_{\mu\in \cal{S}}T_{\mu\kappa}
\sum_{\cal{P}\subseteq \cal{S}\setminus\{\mu\}}\Bigg\{\langle \hat{X}_\kappa(t)
[\hat{A}_{\cal{P}\cup\{\mu\}}(t),\hat{X}^\dagger_\mu(t)]\hat{B}_{\cal{P}\cup\{\mu\}}(t')\rangle^\mathrm{c}\nonumber\\
& &
+\sum_{\cal{Q}\subseteq \cal{P}}^{\cal{Q}\cup \bar{\cal{Q}}=\cal{P}}
\langle [\hat{A}_{\cal{Q}\cup\{\mu\}}(t),\hat{X}^\dagger_\mu(t)]\hat{B}_{\cal{Q}\cup\{\mu\}}(t')\rangle^\mathrm{c} 
\langle \hat{X}_\kappa(t)\hat{A}_{\bar{\cal{Q}}}(t)\hat{B}_{\bar{\cal{Q}}}(t')\rangle^\mathrm{c}\Bigg\}
\sum_{\cup_i{\cal{P}}_i=\cal{S}\setminus\cal{P}\cup\{\mu\}}
\prod_{{\cal{P}}_i}\langle \hat{A}_{{\cal{P}}_i}(t)\hat{B}_{{\cal{P}}_i}(t')\rangle^\mathrm{c}
\end{eqnarray}
Now we consider the equations of motion for the product of all partitions which contain
\textit{proper} subsets of $\cal{S}$.
Applying the product rule of differentiation, we find
\begin{eqnarray}\label{parts}
i\partial_t \sum_{\cup_i {\cal{P}}_i={\cal{S}},{{\cal P}}_i\subset {\cal{S}}} \prod_{{\cal P}_i}\langle \hat{A}_{{\cal{P}}_i}(t)\hat{B}_{{\cal{P}}_i}(t')\rangle^\mathrm{c}=
\sum_{{\cal P}\subset{\cal S}}(i\partial_t\langle \hat{A}_{{\cal{P}}}(t)\hat{B}_{{\cal{P}}}(t')\rangle^\mathrm{c})
\sum_{\cup_i {\cal{P}}_i={\cal{S}}\setminus {\cal P}} \prod_{{\cal P}_i}\langle \hat{A}_{{\cal{P}}_i}(t)\hat{B}_{{\cal{P}}_i}(t')\rangle^\mathrm{c}\,.
\end{eqnarray}
The time derivative of $\langle \hat{A}_{{\cal{P}}}(t)\hat{B}_{{\cal{P}}}(t')\rangle^\mathrm{c}$ can be expressed using
(\ref{hierarchy1})-(\ref{hierarchy9}).
Substituting the first term in (\ref{hierarchy1}) to equation (\ref{parts}) leads to
\begin{eqnarray}
(0')\equiv\sum_{{\cal P}\subset {\cal S}\setminus\{\mu\}}\langle [\hat{A}_{\mathcal{P}\cup\{\mu\}}(t),\hat{H}_\mu(t)]\hat{B}_{\mathcal{P}\cup\{\mu\}}(t')\rangle
\sum_{\cup_i{\cal{P}}_i=\cal{S}\setminus\cal{P}\cup\{\mu\}}
\prod_{{\cal{P}}_i}\langle \hat{A}_{{\cal{P}}_i}(t)\hat{B}_{{\cal{P}}_i}(t')\rangle^\mathrm{c}\,,
\end{eqnarray}
which is the expression (\ref{zero}) without the summand ${\cal P}={\cal S}\setminus\{\mu\}$.
Therefore we find
\begin{eqnarray}\label{h1}
(0)-(0')= \sum_{\mu \in {\cal S}} \langle [\hat{A}_{\mathcal{S}}(t),\hat{H}_\mu(t)]\hat{B}_{\cal S}(t')\rangle^\mathrm{c}\,.
\end{eqnarray}
The contribution of the second term in (\ref{hierarchy1}) to equation (\ref{parts}) gives
\begin{eqnarray}\label{partoneprime}
(1')\equiv -\frac{J}{Z}\sum_{{\cal P}\subset{\cal S}}\sum_{\mu,\nu\in{{\cal P}}}T_{\mu\nu}
\langle [\hat{A}_{{\cal P}}(t),\hat{X}_\mu^\dagger(t)\hat{X}_\nu(t)]\hat{B}_{{\cal P}}(t')\rangle^\mathrm{c}
\sum_{\cup_i {\cal{P}}_i={\cal{S}}\setminus {\cal P}} \prod_{{\cal P}_i}\langle \hat{A}_{{\cal{P}}_i}(t)\hat{B}_{{\cal{P}}_i}(t')\rangle^\mathrm{c}
\end{eqnarray}
In equation (\ref{partoneprime}), the term ${\cal P}={\cal S}$ is excluded from the summation, contrary to equation (\ref{first}), thus
\begin{eqnarray}\label{h2}
(1)-(1')=-\frac{J}{Z}\sum_{\mu,\nu\in\mathcal{S}}T_{\mu\nu}\langle 
[\hat{A}_{\mathcal{S}}(t),\hat{X}_\mu^\dagger(t)\hat{X}_\nu(t)]\hat{B}_{\cal S}(t')\rangle^\mathrm{c}\,.
 \end{eqnarray}
The term (\ref{hierarchy5}) together with (\ref{parts}) leads to
\begin{eqnarray}\label{parttwoaprime}
(2a')&\equiv& -\frac{J}{Z}\sum_{{\cal P}\subset{\cal S}}\sum_{\mu,\nu\in{{\cal P}}}\sum_{{\cal Q}\subseteq {\cal P}\setminus\{\mu,\nu\}}
^{{\cal Q}\cup \bar{{\cal Q}}={\cal P}\setminus\{\mu,\nu\}}T_{\mu\nu}
\langle[\hat{A}_{\cal{Q}\cup\{\mu\}}(t),\hat{X}^\dagger_\mu(t)]\hat{B}_{\cal{Q}\cup\{\mu\}}(t')\rangle^\mathrm{c} 
\langle \hat{X}_\nu(t)\hat{A}_{\bar{\cal{Q}}\cup\{\nu\}}(t)\hat{B}_{\bar{\cal{Q}}\cup\{\nu\}}(t')\rangle^\mathrm{c}
\nonumber\\
& &\times\sum_{\cup_i {\cal{P}}_i={\cal{S}}\setminus {\cal P}} \prod_{{\cal P}_i}\langle \hat{A}_{{\cal{P}}_i}(t)\hat{B}_{{\cal{P}}_i}(t')\rangle^\mathrm{c}
\end{eqnarray}
Using the same reasoning as above we find the difference
\begin{eqnarray}\label{h3}
(2a)-(2a')=-\frac{J}{Z}\sum_{\mu,\nu\in{{\cal S}}}\sum_{{\cal Q}\subseteq {\cal S}\setminus\{\mu,\nu\}}
^{{\cal Q}\cup \bar{{\cal Q}}={\cal S}\setminus\{\mu,\nu\}}
T_{\mu\nu}\langle[\hat{A}_{\cal{Q}\cup\{\mu\}}(t),\hat{X}^\dagger_\mu(t)]\hat{B}_{\cal{Q}\cup\{\mu\}}(t')\rangle^\mathrm{c} 
\langle \hat{X}_\nu(t)\hat{A}_{\bar{\cal{Q}}\cup\{\nu\}}(t)\hat{B}_{\bar{\cal{Q}}\cup\{\nu\}}(t')\rangle^\mathrm{c} 
\end{eqnarray}
The part (\ref{hierarchy6}) from the hierarchy together with (\ref{parts}) leads to an expression $(2b')$
which has to be substracted from $(2b)$ from equation (\ref{first}),
\begin{eqnarray}\label{h4}
(2b)-(2b')=-\frac{J}{Z}\sum_{\mu,\nu\in{{\cal S}}}T_{\mu\nu}\sum_{{\cal Q}\subseteq {\cal S}\setminus\{\mu,\nu\}}
^{{\cal Q}\cup \bar{{\cal Q}}={\cal S}\setminus\{\mu,\nu\}}
\langle \hat{A}_{\cal{Q}\cup\{\mu\}}(t)\hat{X}^\dagger_\mu(t)\hat{B}_{\cal{Q}\cup\{\mu\}}(t')\rangle^\mathrm{c} 
\langle [\hat{A}_{\bar{\cal{Q}}\cup\{\nu\}}(t),\hat{X}_\nu(t)]\hat{B}_{\bar{\cal{Q}}\cup\{\nu\}}(t')\rangle^\mathrm{c}\,.
\end{eqnarray}
The contribution of the first term in (\ref{hierarchy2}) to equation (\ref{parts}) leads to
\begin{eqnarray}
(3a')\equiv-\frac{J}{Z}\sum_{{\cal P}\subset {\cal S}}\sum_{\kappa \notin {\cal P}} \sum_{\mu\in {\cal P}}T_{\mu\kappa}
\langle \hat{X}_\kappa^\dagger(t)[\hat{A}_{\cal P}(t),\hat{X}_\mu(t)]\hat{B}_{\cal P}(t')\rangle^\mathrm{c}\sum_{\cup_i {\cal{P}}_i={\cal{S}}\setminus 
{\cal P}} \prod_{{\cal P}_i}\langle \hat{A}_{{\cal{P}}_i}(t)\hat{B}_{{\cal{P}}_i}(t')\rangle^\mathrm{c}
\end{eqnarray}
whereas the first term of (\ref{hierarchy7}) together with (\ref{parts}) gives
\begin{eqnarray}
(3a'')&\equiv&\frac{J}{Z}\sum_{{\cal P}\subset {\cal S}}\sum_{\mu,\nu\in {\cal P}}\sum_{{\cal Q}\subseteq {\cal P}\setminus \{\mu,\nu\}}
^{{\cal Q}\cup \bar{{\cal Q}}={\cal P}\setminus \{\mu,\nu\}}T_{\mu\nu}\langle \hat{A}_{{\cal Q}\cup \{ \nu\}}(t)\hat{B}_{{\cal Q}\cup \{ \nu\}}(t')\rangle^\mathrm{c}
\langle \hat{X}^\dagger_\nu(t)[\hat{A}_{\bar{\cal Q}\cup \{\mu\}}(t),\hat{X}_\mu(t)]\hat{B}_{\bar{\cal Q}\cup \{\mu\}}(t')\rangle^\mathrm{c}\nonumber\\
& &\times \sum_{\cup_i {\cal{P}}_i={\cal{S}}\setminus 
{\cal P}} \prod_{{\cal P}_i}\langle \hat{A}_{{\cal{P}}_i}(t)\hat{B}_{{\cal{P}}_i}(t')\rangle^\mathrm{c}
\end{eqnarray}
Splitting the sum over lattice sites $\kappa$ according to $\sum_{\kappa \neq {\cal P}}=\sum_{\kappa\notin {\cal S}}+\sum_{\kappa\in {\cal S}\setminus{\cal S}}$
and substracting $(3a')$ as well as $(3a'')$ from $(3a)$ in (\ref{second}) leads to
\begin{eqnarray}\label{h5}
(3a)-(3a')-(3a'')&=&-\frac{J}{Z}\sum_{\kappa \notin {\cal S}}\sum_{\mu\in {\cal S}}T_{\mu\kappa}
\langle \hat{X}^\dagger_\kappa(t)[\hat{A}_{{\cal S}}(t),\hat{X}_\mu(t)]\hat{B}_{{\cal S}}(t')\rangle^\mathrm{c}\nonumber
\\& &+\frac{J}{Z}\sum_{\mu,\nu \in {\cal S}}
\sum_{{\cal P}\subseteq {\cal S}\setminus \{\mu,\nu\}}
^{{\cal P}\cup \bar{{\cal P}}={\cal S}\setminus \{\mu,\nu\}} T_{\mu\nu}
\langle \hat{A}_{{\cal P}\cup \{ \nu\}}(t)\hat{B}_{{\cal P}\cup \{ \nu\}}(t')\rangle^\mathrm{c}
\langle \hat{X}_\nu^\dagger(t)[\hat{A}_{\bar {\cal P}\cup \{\mu\}}(t),\hat{X}_\mu(t)]\hat{B}_{\bar {\cal P}\cup \{\mu\}}(t')\rangle^\mathrm{c}
\end{eqnarray}
Similar terms $(3b')$ and $(3b'')$ derive from the first terms in (\ref{hierarchy2}) and (\ref{hierarchy6}) together with (\ref{parts}).
Substracting them from $(3b)$ which was given in relation (\ref{third}) leads to
\begin{eqnarray}\label{h6}
(3b)-(3b')-(3b'')&=&-\frac{J}{Z}\sum_{\kappa \notin {\cal S}}\sum_{\mu\in {\cal S}}T_{\mu\kappa}
\langle \hat{X}_\kappa(t)[\hat{A}_{{\cal S}}(t),\hat{X}^\dagger_\mu(t)]\hat{B}_{{\cal S}}(t')\rangle^\mathrm{c}\nonumber
\\& &+\frac{J}{Z}\sum_{\mu,\nu \in {\cal S}}
\sum_{{\cal P}\subseteq {\cal S}\setminus \{\mu,\nu\}}
^{{\cal P}\cup \bar{{\cal P}}={\cal S}\setminus \{\mu,\nu\}} T_{\mu\nu}
\langle \hat{A}_{{\cal P}\cup \{ \nu\}}(t)\hat{B}_{{\cal P}\cup \{ \nu\}}(t')\rangle^\mathrm{c}
\langle \hat{X}_\nu(t)[\hat{A}_{\bar {\cal P}\cup \{\mu\}}(t),\hat{X}^\dagger_\mu(t)]\hat{B}_{\bar {\cal P}\cup \{\mu\}}(t')\rangle^\mathrm{c}
\end{eqnarray}
The contribution of (\ref{hierarchy3}) in relation (\ref{parts}) results in 
\begin{eqnarray}
(4a')&\equiv&-\frac{J}{Z}\sum_{{\cal P}\subset{\cal S}}\sum_{\kappa \notin{{\cal P}}}\sum_{\mu\in {\cal P}}
\sum_{{\cal Q}\subseteq {\cal P}\setminus \{\mu\}}^{{\cal Q}\cup\bar{{\cal Q}}={\cal P}\setminus\{\mu\}}
T_{\mu\kappa}\langle [\hat{A}_{{\cal Q}\cup \{\mu\}}(t),\hat{X}_\mu(t)]\hat{B}_{{\cal Q}\cup \{\mu\}}(t')\rangle^\mathrm{c} 
\langle \hat{X}_\kappa^\dagger(t)\hat{A}_{\bar{{\cal Q}}}(t)\hat{B}_{\bar{{\cal Q}}}(t')\rangle^\mathrm{c}
\nonumber\\
& &\times\sum_{\cup_i {\cal{P}}_i={\cal{S}}\setminus 
{\cal P}} \prod_{{\cal P}_i}\langle \hat{A}_{{\cal{P}}_i}(t)\hat{B}_{{\cal{P}}_i}(t')\rangle^\mathrm{c}
\end{eqnarray}
whereas (\ref{hierarchy8}) together with (\ref{parts}) gives
\begin{eqnarray}
(4a'')&\equiv&\frac{J}{Z}\sum_{{\cal P}\subset {\cal S}}\sum_{\mu,\nu\in {\cal P}}
\sum_{{\cal R}\subseteq {\cal P}\setminus\{\mu,\nu\}}^{{\cal R}\cup \bar{{\cal R}}={\cal P}\setminus \{\mu,\nu\}}
T_{\mu\nu}\langle \hat{A}_{{\cal R}\cup \{\nu\}}(t)\hat{B}_{{\cal R}\cup \{\nu\}}(t')\rangle^\mathrm{c}\sum_{{\cal Q}\subseteq \bar{{\cal R}}}^{{\cal Q}\cup
\bar{{\cal Q}}=\bar{{\cal R}}}\langle [\hat{A}_{{\cal Q}\cup\{\mu\}}(t),\hat{X}_\mu(t)]\hat{B}_{{\cal Q}\cup\{\mu\}}(t')\rangle^\mathrm{c}
\langle \hat{A}_{\bar{{\cal Q}}}(t)\hat{X}_\nu(t)\hat{B}_{\bar{{\cal Q}}}(t')\rangle^\mathrm{c}\nonumber\\
& &\times\sum_{\cup_i {\cal{P}}_i={\cal{S}}\setminus 
{\cal P}} \prod_{{\cal P}_i}\langle \hat{A}_{{\cal{P}}_i}(t)\hat{B}_{{\cal{P}}_i}(t')\rangle^\mathrm{c}\,.
\end{eqnarray}
Splitting the sum over $\kappa$ as above and substracting
$(4a')$ and $(4a'')$ from $(4a)$ which was defined in relation (\ref{third}) leads to
\begin{eqnarray}\label{h7}
(4a)-(4a')-(4a'')=-\frac{J}{Z}\sum_{\kappa \notin {\cal S}}\sum_{\mu \in {\cal S}}T_{\mu\kappa}
\sum_{{\cal P}\subseteq {\cal S}\setminus\{\mu\}}^{{\cal P}\cup\bar{{\cal P}}={\cal S}\setminus\{\mu\}}
\langle [\hat{A}_{{\cal P}\cup \{\mu\}}(t),\hat{X}_\mu(t)]\hat{B}_{{\cal P}\cup \{\mu\}}(t')\rangle^\mathrm{c}
\langle \hat{X}_\kappa^\dagger (t)\hat{A}_{\bar{{\cal P}}}(t)\hat{B}_{\bar{{\cal P}}}(t')\rangle^\mathrm{c}\nonumber\\
+\frac{J}{Z}\sum_{\mu\nu}
\sum_{{\cal P}\subseteq {\cal S}\setminus\{\mu,\nu\}}^{{\cal P}\cup \bar{{\cal P}}={\cal S}\setminus\{\mu,\nu\}}
T_{\mu\nu}\sum_{{\cal Q}\subseteq {\bar{{\cal P}}}}^{{\cal Q}\cup \bar{{\cal Q}}=\bar{{\cal P}}} 
\langle [\hat{A}_{{\cal Q}\cup \{\mu\}}(t),\hat{X}_\mu(t)]\hat{B}_{{\cal Q}\cup \{\mu\}}(t')\rangle^\mathrm{c}
\langle \hat{X}^\dagger_\nu(t)\hat{A}_{\bar{{\cal Q}}}(t)\hat{B}_{\bar{{\cal Q}}}(t')\rangle^\mathrm{c}
\langle \hat{A}_{{\cal P}\cup \{\nu\}}(t)\hat{B}_{{\cal P}\cup \{\nu\}}(t')\rangle^\mathrm{c}
\end{eqnarray}
The terms $(4b')$ and $(4b'')$ similar $(4a')$ and $(4a'')$ can be 
derived from (\ref{hierarchy4}) and (\ref{hierarchy9}).
After substracting them from $(4b)$ in relation (\ref{third}) we have
\begin{eqnarray}\label{h8}
(4b)-(4b')-(4b'')=-\frac{J}{Z}\sum_{\kappa \notin {\cal S}}\sum_{\mu \in {\cal S}}T_{\mu\kappa}
\sum_{{\cal P}\subseteq {\cal S}\setminus\{\mu\}}^{{\cal P}\cup\bar{{\cal P}}={\cal S}\setminus\{\mu\}}
\langle [\hat{A}_{{\cal P}\cup \{\mu\}}(t),\hat{X}^\dagger_\mu(t)]\hat{B}_{{\cal P}\cup \{\mu\}}(t')\rangle^\mathrm{c}
\langle \hat{X}_\kappa(t)\hat{A}_{\bar{{\cal P}}}(t)\hat{B}_{\bar{{\cal P}}}(t')\rangle^\mathrm{c}\nonumber\\
+\frac{J}{Z}\sum_{\mu\nu}
\sum_{{\cal P}\subseteq {\cal S}\setminus\{\mu,\nu\}}^{{\cal P}\cup \bar{{\cal P}}={\cal S}\setminus\{\mu,\nu\}}
T_{\mu\nu}\sum_{{\cal Q}\subseteq {\bar{{\cal P}}}}^{{\cal Q}\cup \bar{{\cal Q}}=\bar{{\cal P}}} 
\langle [\hat{A}_{{\cal Q}\cup \{\mu\}}(t),\hat{X}^\dagger_\mu(t)]\hat{B}_{{\cal Q}\cup \{\mu\}}(t')\rangle^\mathrm{c}
\langle \hat{X}_\nu(t)\hat{A}_{\bar{{\cal Q}}}(t)\hat{B}_{\bar{{\cal Q}}}(t')\rangle^\mathrm{c}
\langle \hat{A}_{{\cal P}\cup \{\nu\}}(t)\hat{B}_{{\cal P}\cup \{\nu\}}(t')\rangle^\mathrm{c}\,.
\end{eqnarray}
Adding the equations (\ref{h1}), (\ref{h2}), (\ref{h3}), (\ref{h4}), (\ref{h5}), (\ref{h6}), 
(\ref{h7}) and (\ref{h8}) together leads to the right hand side of the hierarchy (\ref{hierarchy1})-(\ref{hierarchy9})
for the a set with cardinality $|{\cal S}|$.
This completes the proof.

%

\section{appendix: double-time correlation functions}\label{doubtime}
In the following, we choose the abbreviation $J^\pm_\mathbf{k}=J^\pm T_\mathbf{k}$, $t^+=(t+t')/2$
and $t^-=(t-t')/2$.
For $t,t'<0$, the Fourier components of the correlation 
functions read
\begin{align}
f^{11}_\mathbf{k}&= 
\frac{U-3J^-_\mathbf{k}+\omega_\mathbf{k}^-}{2\omega_\mathbf{k}^-}e^{\frac{i(t-t')}{2}\left(U-J^-_\mathbf{k}-\omega_\mathbf{k}^-\right)}-1\\
f^{22}_\mathbf{k}&= 
\frac{U-3J^-_\mathbf{k}-\omega_\mathbf{k}^-}{2\omega_\mathbf{k}^-}e^{\frac{i(t-t')}{2}\left(U-J^-_\mathbf{k}-\omega_\mathbf{k}^-\right)}\\
f^{12}_\mathbf{k}&=f^{21}_\mathbf{k}=
\frac{\sqrt{2}J^-_\mathbf{k}}{\omega_\mathbf{k}^-}e^{\frac{i(t-t')}{2}\left(U-J^-_\mathbf{k}-\omega_\mathbf{k}^-\right)}\,,
\end{align}
and the light-cone structure in $D$ dimensions is determined by $\pm 2 J^- t^-/D=x$ along
a lattice axis and $\pm 2 J^- t^-/\sqrt{D}=x$ along a diagonal.
For $t>0$ and $t'<0$, we have
\begin{align}
f^{11}_\mathbf{k}=&-1-\frac{e^{\frac{i}{2}[t(U+\omega^+_\mathbf{k}-J^+_\mathbf{k})-t'(U-\omega_\mathbf{k}^--J^-_\mathbf{k})]}}
{4\omega^+_\mathbf{k}\omega^-_\mathbf{k}}\left[(U-3 J^- _\mathbf{k}+\omega^-_\mathbf{k})(U-3 J^+ _\mathbf{k}-\omega^+_\mathbf{k})-8J^+ _\mathbf{k}J^- _\mathbf{k}\right]\nonumber\\
&+\frac{e^{\frac{i}{2}[t(U-\omega^+_\mathbf{k}-J^+_\mathbf{k})-t'(U-\omega_\mathbf{k}^--J^-_\mathbf{k})]}}
{4\omega^+_\mathbf{k}\omega^-_\mathbf{k}}\left[(U-3 J^- _\mathbf{k}+\omega^-_\mathbf{k})(U-3 J^+ _\mathbf{k}+\omega^+_\mathbf{k})-8J^+ _\mathbf{k}J^- _\mathbf{k}\right]\\
f^{12}_\mathbf{k}=&\frac{e^{\frac{i}{2}[t(U+\omega^+_\mathbf{k}-J^+_\mathbf{k})-t'(U-\omega_\mathbf{k}^--J^-_\mathbf{k})]}}
{\sqrt{2}\omega^+_\mathbf{k}\omega^-_\mathbf{k}}
[J^+_\mathbf{k}(U-\omega^-_\mathbf{k})-J^-_\mathbf{k}(U-\omega^+_\mathbf{k})]\nonumber\\
& +\frac{e^{\frac{i}{2}[t(U-\omega^+_\mathbf{k}-J^+_\mathbf{k})-t'(U-\omega_\mathbf{k}^--J^-_\mathbf{k})]}}
{\sqrt{2}\omega^+_\mathbf{k}\omega^-_\mathbf{k}}
[J^-_\mathbf{k}(U+\omega^+_\mathbf{k})-J^+_\mathbf{k}(U-\omega^-_\mathbf{k})]\\
f^{21}_\mathbf{k}=&\frac{e^{\frac{i}{2}[t(U+\omega^+_\mathbf{k}-J^+_\mathbf{k})-t'(U-\omega_\mathbf{k}^--J^-_\mathbf{k})]}}
{\sqrt{2}\omega^+_\mathbf{k}\omega^-_\mathbf{k}}
[J^-_\mathbf{k}(U+\omega^+_\mathbf{k})-J^+_\mathbf{k}(U+\omega^-_\mathbf{k})]\nonumber\\
& +\frac{e^{\frac{i}{2}[t(U-\omega^+_\mathbf{k}-J^+_\mathbf{k})-t'(U-\omega_\mathbf{k}^--J^-_\mathbf{k})]}}
{\sqrt{2}\omega^+_\mathbf{k}\omega^-_\mathbf{k}}
[J^+_\mathbf{k}(U+\omega^-_\mathbf{k})-J^-_\mathbf{k}(U-\omega^+_\mathbf{k})]\\
f^{22}_\mathbf{k}=&\frac{e^{\frac{i}{2}[t(U+\omega^+_\mathbf{k}-J^+_\mathbf{k})-t'(U-\omega_\mathbf{k}^--J^-_\mathbf{k})]}}
{4\omega^+_\mathbf{k}\omega^-_\mathbf{k}}\left[(U-3 J^- _\mathbf{k}-\omega^-_\mathbf{k})(U-3 J^+ _\mathbf{k}+\omega^+_\mathbf{k})-8J^+ _\mathbf{k}J^- _\mathbf{k}\right]\nonumber\\
&-\frac{e^{\frac{i}{2}[t(U-\omega^+_\mathbf{k}-J^+_\mathbf{k})-t'(U-\omega_\mathbf{k}^--J^-_\mathbf{k})]}}
{4\omega^+_\mathbf{k}\omega^-_\mathbf{k}}\left[(U-3 J^- _\mathbf{k}-\omega^-_\mathbf{k})(U-3 J^+ _\mathbf{k}-\omega^+_\mathbf{k})-8J^+ _\mathbf{k}J^- _\mathbf{k}\right]\,.
\end{align}
Now there are two light-cone structures along the lattice axes: $\pm[(2J^+-J^-)t^-+(2J^++J^-)t^+]/D=x$ and
$\pm[(J^++J^-)t^-+(J^+-J^-)t^+]/D=x$.
Along the diagonals we have 
$\pm[(2J^+-J^-)t^-+(2J^++J^-)t^+]/\sqrt{D}=x$ and
$\pm[(J^++J^-)t^-+(J^+-J^-)t^+]/\sqrt{D}=x$.
For $t<0$ and $t'>0$, the Fourier coefficients read
\begin{align}
f^{11}_\mathbf{k}=&-1+\frac{e^{\frac{i}{2}[t(U-\omega^+_\mathbf{k}-J^+_\mathbf{k})-t'(U-\omega_\mathbf{k}^--J^-_\mathbf{k})]}}
{4\omega^+_\mathbf{k}\omega^-_\mathbf{k}}\left[(U-3 J^- _\mathbf{k}+\omega^-_\mathbf{k})(U-3 J^+ _\mathbf{k}+\omega^+_\mathbf{k})-8J^+ _\mathbf{k}J^- _\mathbf{k}\right]\nonumber\\
&-\frac{e^{\frac{i}{2}[t(U-\omega^+_\mathbf{k}-J^+_\mathbf{k})-t'(U+\omega_\mathbf{k}^--J^-_\mathbf{k})]}}
{4\omega^+_\mathbf{k}\omega^-_\mathbf{k}}\left[(U-3 J^- _\mathbf{k}+\omega^-_\mathbf{k})(U-3 J^+ _\mathbf{k}-\omega^+_\mathbf{k})-8J^+ _\mathbf{k}J^- _\mathbf{k}\right]\\
f^{12}_\mathbf{k}=&\frac{e^{\frac{i}{2}[t(U-\omega^+_\mathbf{k}-J^+_\mathbf{k})-t'(U-\omega_\mathbf{k}^--J^-_\mathbf{k})]}}
{\sqrt{2}\omega^+_\mathbf{k}\omega^-_\mathbf{k}}
[J^+_\mathbf{k}(U+\omega^-_\mathbf{k})-J^-_\mathbf{k}(U-\omega^+_\mathbf{k})]\nonumber\\
& +\frac{e^{\frac{i}{2}[t(U-\omega^+_\mathbf{k}-J^+_\mathbf{k})-t'(U+\omega_\mathbf{k}^--J^-_\mathbf{k})]}}
{\sqrt{2}\omega^+_\mathbf{k}\omega^-_\mathbf{k}}
[J^-_\mathbf{k}(U+\omega^+_\mathbf{k})-J^+_\mathbf{k}(U+\omega^-_\mathbf{k})]\\
f^{21}_\mathbf{k}=&\frac{e^{\frac{i}{2}[t(U+\omega^+_\mathbf{k}-J^+_\mathbf{k})-t'(U-\omega_\mathbf{k}^--J^-_\mathbf{k})]}}
{\sqrt{2}\omega^+_\mathbf{k}\omega^-_\mathbf{k}}
[J^-_\mathbf{k}(U+\omega^+_\mathbf{k})-J^+_\mathbf{k}(U-\omega^-_\mathbf{k})]\nonumber\\
& +\frac{e^{\frac{i}{2}[t(U-\omega^+_\mathbf{k}-J^+_\mathbf{k})-t'(U-\omega_\mathbf{k}^--J^-_\mathbf{k})]}}
{\sqrt{2}\omega^+_\mathbf{k}\omega^-_\mathbf{k}}
[J^+_\mathbf{k}(U-\omega^-_\mathbf{k})-J^-_\mathbf{k}(U-\omega^+_\mathbf{k})]\\
f^{22}_\mathbf{k}=&-\frac{e^{\frac{i}{2}[t(U+\omega^+_\mathbf{k}-J^+_\mathbf{k})-t'(U-\omega_\mathbf{k}^--J^-_\mathbf{k})]}}
{4\omega^+_\mathbf{k}\omega^-_\mathbf{k}}\left[(U-3 J^- _\mathbf{k}-\omega^-_\mathbf{k})(U-3 J^+ _\mathbf{k}-\omega^+_\mathbf{k})-8J^+ _\mathbf{k}J^- _\mathbf{k}\right]\nonumber\\
&+\frac{e^{\frac{i}{2}[t(U-\omega^+_\mathbf{k}-J^+_\mathbf{k})-t'(U-\omega_\mathbf{k}^--J^-_\mathbf{k})]}}
{4\omega^+_\mathbf{k}\omega^-_\mathbf{k}}\left[(U-3 J^- _\mathbf{k}-\omega^-_\mathbf{k})(U-3 J^+ _\mathbf{k}+\omega^+_\mathbf{k})-8J^+ _\mathbf{k}J^- _\mathbf{k}\right]\,,
\end{align}
Also here are two light-cone structures along the lattice axes: 
\begin{align}
\pm[(2J^--J^+)t^--(2J^-+J^+)t^+]/D&=x\\
\pm[(J^++J^-)t^-+(J^+-J^-)t^+]/D&=x\,.
\end{align}
Along the diagonals we have 
\begin{align}
\pm[(2J^--J^+)t^--(2J^-+J^+)t^+]/\sqrt{D}&=x\\
\pm[(J^++J^-)t^-+(J^+-J^-)t^+]/\sqrt{D}&=x\,.
\end{align}
and finally we have for $t,t'>0$
\begin{align}
f^{11}_\mathbf{k}=&-1+\frac{e^{\frac{i}{2}(t-t')(U+\omega^+_\mathbf{k}-J^+_\mathbf{k})}}
{4\omega^-_\mathbf{k}(\omega^+_\mathbf{k})^2}\nonumber\\
&\times[8 J^+_\mathbf{k}U(J^+_\mathbf{k}-J^-_\mathbf{k})
+3\omega^+_\mathbf{k}(J^+_\mathbf{k}(U+\omega^-_\mathbf{k})+J^-_\mathbf{k}(U-\omega^+_\mathbf{k}))
-\omega^+_\mathbf{k}(J^+_\mathbf{k}J^-_\mathbf{k}+(U+\omega^-_\mathbf{k})(U-\omega^+_\mathbf{k})) ]\nonumber\\
+&\frac{e^{\frac{i}{2}(t-t')(U-\omega^+_\mathbf{k}-J^+_\mathbf{k})}}
{4\omega^-_\mathbf{k}(\omega^+_\mathbf{k})^2}\nonumber\\
&\times[8 J^+_\mathbf{k}U(J^+_\mathbf{k}-J^-_\mathbf{k})
-3\omega^+_\mathbf{k}(J^+_\mathbf{k}(U+\omega^-_\mathbf{k})+J^-_\mathbf{k}(U+\omega^+_\mathbf{k}))
+\omega^+_\mathbf{k}(J^+_\mathbf{k}J^-_\mathbf{k}+(U+\omega^-_\mathbf{k})(U+\omega^+_\mathbf{k})) ]\nonumber\\
+&\left(e^{\frac{i}{2}[t(U-\omega^+_\mathbf{k}-J^+_\mathbf{k})-t'(U+\omega^+_\mathbf{k}-J^+_\mathbf{k})]}+
e^{\frac{i}{2}[t(U+\omega^+_\mathbf{k}-J^+_\mathbf{k})-t'(U-\omega^+_\mathbf{k}-J^+_\mathbf{k})]}\right)
\frac{2J^+_\mathbf{k}U(J^-_\mathbf{k}-J^+_\mathbf{k})} {\omega^-_\mathbf{k}(\omega^+_\mathbf{k})^2}\\
f^{12}_\mathbf{k}=f^{21}_\mathbf{k}=&\frac{e^{\frac{i}{2}(t-t')(U+\omega^+_\mathbf{k}-J^+_\mathbf{k})}}
{\sqrt{2}\omega^-_\mathbf{k}(\omega^+_\mathbf{k})^2}J^+_\mathbf{k}[J^-_\mathbf{k}J^+_\mathbf{k}-3 U(J^+_\mathbf{k}+J^-_\mathbf{k})+
U^2-\omega^+_\mathbf{k}\omega^-_\mathbf{k}]\nonumber\\
+&\frac{e^{\frac{i}{2}(t-t')(U-\omega^+_\mathbf{k}-J^+_\mathbf{k})}}
{\sqrt{2}\omega^-_\mathbf{k}(\omega^+_\mathbf{k})^2}J^+_\mathbf{k}[J^-_\mathbf{k}J^+_\mathbf{k}-3 U(J^+_\mathbf{k}+J^-_\mathbf{k})+
U^2+\omega^+_\mathbf{k}\omega^-_\mathbf{k}]\nonumber\\
+&\frac{e^{\frac{i}{2}[t(U-\omega^+_\mathbf{k}-J^+_\mathbf{k})-t'(U+\omega^+_\mathbf{k}-J^+_\mathbf{k})]}}{\sqrt{2}\omega^-_\mathbf{k}(\omega^+_\mathbf{k})^2}
(J^-_\mathbf{k}-J^+_\mathbf{k})U(U+\omega^+_\mathbf{k}-3J^+_\mathbf{k})\nonumber\\
+&\frac{e^{\frac{i}{2}[t(U+\omega^+_\mathbf{k}-J^+_\mathbf{k})-t'(U-\omega^+_\mathbf{k}-J^+_\mathbf{k})]}}{\sqrt{2}\omega^-_\mathbf{k}(\omega^+_\mathbf{k})^2}
(J^-_\mathbf{k}-J^+_\mathbf{k})U(U-\omega^+_\mathbf{k}-3J^+_\mathbf{k})\\
f^{22}_\mathbf{k}=&\frac{e^{\frac{i}{2}(t-t')(U+\omega^+_\mathbf{k}-J^+_\mathbf{k})}}
{8\omega^-_\mathbf{k}(\omega^+_\mathbf{k})^2}\nonumber\\
&\times[8 J^+_\mathbf{k}(J^+_\mathbf{k}(U+\omega^-_\mathbf{k})-J^-_\mathbf{k}(U+\omega^+_\mathbf{k}))
+(U+\omega^+_\mathbf{k}-3 J^+_\mathbf{k})[(U-\omega^-_\mathbf{k})(\omega^+_\mathbf{k}+U-3 J^+_\mathbf{k})+J^-_\mathbf{k}
(J^+_\mathbf{k}-3(U+\omega^+_\mathbf{k})) ]]
\nonumber\\
+&\frac{e^{\frac{i}{2}(t-t')(U-\omega^+_\mathbf{k}-J^+_\mathbf{k})}}
{8\omega^-_\mathbf{k}(\omega^+_\mathbf{k})^2}\nonumber\\
&\times[8 J^+_\mathbf{k}(J^+_\mathbf{k}(U+\omega^-_\mathbf{k})-J^-_\mathbf{k}(U-\omega^+_\mathbf{k}))
-(U-\omega^+_\mathbf{k}-3 J^+_\mathbf{k})[(U-\omega^-_\mathbf{k})(\omega^+_\mathbf{k}-U+3 J^+_\mathbf{k})-J^-_\mathbf{k}
(J^+_\mathbf{k}-3(U-\omega^+_\mathbf{k})) ]]
\nonumber\\
+&\left(e^{\frac{i}{2}[t(U-\omega^+_\mathbf{k}-J^+_\mathbf{k})-t'(U+\omega^+_\mathbf{k}-J^+_\mathbf{k})]}
+e^{\frac{i}{2}[t(U+\omega^+_\mathbf{k}-J^+_\mathbf{k})-t'(U-\omega^+_\mathbf{k}-J^+_\mathbf{k})]}\right)
\frac{2J^+_\mathbf{k}U(J^-_\mathbf{k}-J^+_\mathbf{k})}{\omega^-_\mathbf{k}(\omega^+_\mathbf{k})^2}
\end{align}
Here we have four light-cones which are determined 
by
\begin{align}
\pm 4 J^+ t^-/D&=x\\
\pm 2 J^+ t^-/D&=x\\
\pm(J^+ t^--3 J^+ t^+)/D&=x\\
\pm(J^+ t^-+3 J^+ t^+)/D&=x
\end{align}
and, similarly along the diagonals 
\begin{align}
\pm 4 J^+ t^-/\sqrt{D}&=x\\
\pm 2 J^+ t^-/\sqrt{D}&=x\\
\pm(J^+ t^--3 J^+ t^+)/\sqrt{D}&=x\\
\pm(J^+ t^-+3 J^+ t^+)/\sqrt{D}&=x
\end{align}

\end{widetext}

\end{document}